\documentclass[a4paper,12pt]{article}
\usepackage{amssymb}
\usepackage{amsmath}
\usepackage{amstext}
\usepackage{amsfonts}
\usepackage{epsfig}
\usepackage{dcolumn}
\usepackage{rotating}
\usepackage{color}
\usepackage{comment}
\usepackage{hyperref}

\newcommand{\bbox}{\bar{\Box}}
\newcommand{\bn}{\bold{n}}

\newcommand{\ba}{\begin{eqnarray}}
\newcommand{\ea}{\end{eqnarray}}
\newcommand{\be}{\begin{equation}}
\newcommand{\ee}{\end{equation}}
\newcommand{\bi}{\begin{itemize}}
\newcommand{\ei}{\end{itemize}}
\newcommand{\da}{\delta}
\newcommand{\la}{\lambda}

\newcommand{\sa}{\sigma}

\newcommand{\Ga}{\Gamma}

\newcommand{\cF}{{\cal F}}

\newcommand{\cH}{{\cal H}}

\newcommand{\cL}{{\cal L}}

\newcommand{\cP}{{\cal P}}

\newcommand{\p}{\partial}

\newcommand{\Ra}{\Rightarrow}

\newcommand{\LF}{\left(}
\newcommand{\RF}{\right)}
\newcommand{\LT}{\left[}
\newcommand{\RT}{\right]}

\newcommand{\kb}{\bar{k}}

\newcommand{\non}{\nonumber\\}

\begin{document}

%+Title
\title{Hamiltonian Analysis for Infinite Derivative Field Theory and Gravity}
\author{Spyridon Talaganis, Ali Teimouri \\ \\
 {\it Consortium for Fundamental Physics, Lancaster University,} \\
{\it Lancaster, LA$1$ $4$YB, United Kingdom.}\\
\begin{footnotesize}\textit{E-mail}:  s.talaganis@lancaster.ac.uk, a.teimouri@lancaster.ac.uk\end{footnotesize}}
\date{}

\maketitle
%-Title

%+Abstract
\begin{abstract}
Typically higher-derivative theories are unstable. Instabilities manifest themselves from {\it extra} propagating degrees of freedom, which are unphysical. In this paper, we will investigate an {\it infinite derivative} field theory and study its {\it true} dynamical degrees of freedom via Hamiltonian analysis.
In particular, we will show that if the infinite derivatives can be captured by a Gaussian kinetic term, i.e. {\it exponential of entire function}, then it is possible to prove that there are only {\it finite} number of dynamical degrees of freedom. We will further extend our investigation into infinite derivative theory of gravity, and in particular concentrate on {\it ghost free} and {\it singularity free} theory of gravity, which has been studied extensively in the Lagrangian approach. Here we will show from the Hamiltonian perspective that there are only finite number of degrees of freedom. \end{abstract}
%-Abstract

%+Contents
\tableofcontents
%-Contents

\section{Introduction}
\numberwithin{equation}{section}

It has been known for a while that in four dimensions quadratic curvature gravity is renormalizable~\cite{Stelle:1976gc}. However, being a {\it finite} higher derivative theory of gravity, it contains ghosts, \textit{i.e.} massive spin-2 ghost. This can be seen 
both at a classical and at a quantum level. At a quantum level, one can study the propagator for a graviton in 
a quadratic curvature gravity~\cite{Rivers,VanNieuwenhuizen:1973fi}. This is part of a Lagrangian approach in order to decompose the action in terms of scalar, vector and tensor degrees of freedom, and see whether the action is perturbatively
stable or not. On the other hand, one can as well analyse the Hamiltonian to understand the stability and unboundedness of the 
Hamiltonian density from below, which would typically exhibit Ostr\'ogradsky's instability by virtue of carrying finite number of derivatives higher than two~\cite{Label1}.

Gravity is a diffeomorphism-invariant theory, one would expect all possible diffeomorphism-invariant terms, such as covariant higher- and {\it infinite-derivative}
contributions in the Ricci scalar, Ricci tensor and Weyl~\cite{Biswas:2005qr,Biswas:2011ar,Biswas:2013kla,Biswas:2013cha}. 
Furthermore, a curious observation was made in~\cite{Biswas:2005qr,Biswas:2011ar} that such an infinite derivative action of gravity would not {\it only} address  the {\it ghost} problem of quadratic curvature gravity of Stelle, but would also yields non-singular cosmological solution for homogeneous and isotropic metric~\cite{Biswas:2005qr,Biswas:2010zk,Biswas:2012bp}.  

Note that for such theory every derivative would introduce a new pole in the propagator, and a new degrees of freedom propagating 
in the spacetime. However, graviton being transverse and traceless must carry {\it only} two degrees of freedom in four dimensions, \textit{i.e.} spin-0 and spin-2 components. In particular, it was argued in Refs.~\cite{Biswas:2005qr,Biswas:2011ar,Biswas:2013kla,Biswas:2013cha} that these infinite degrees of freedom can be reduced to the {\it original} $2$ dynamical degrees of freedom provided the propagator gets modified by {\it exponential of an entire function}. Note that similar conclusions regarding ghosts were also being made before in the context of 
IDG in Refs.~\cite{Tomboulis, Tomboulis-1,Tomboulis:2015gfa}, where the author has demanded that the propagator in the UV be modified by an {\it entire function}.

Since, the IDG action contains infinitely many covariant derivatives, there is no highest momentum operator and, as a result, their perturbative stability cannot be analyzed via Ostr\'ogradsky analysis. It begs the question on how to formulate the Hamiltonian for IDG. Indeed, being an infinite derivative theory, the prime questions are - how shall we set the initial conditions, and what are the key dynamical degrees of freedom in this class of theory, what are the {\it primary and secondary constraints}, and what are the {\it first and second-class constraints} as laid down by Dirac~\cite{Anderson:1951ta,Dirac:1958sq,Dirac:1958sc,Dirac1}.

Indeed, seeking the Hamiltonian density for Einstein's gravity is not an easy task, let alone dealing with IDG. However, the 
background works are already very well-known in the literature. In the late 1950s, the $3+1$ decomposition received 
a great deal of attention; Richard Arnowitt, Stanley Deser and Charles W. Misner (ADM) 
\cite{Arnowitt:1962hi, Gourgoulhon:2007ue} have shown that it is possible to decompose four-dimensional spacetime
such that one foliates the arbitrary region $\mathcal{M}$ of the space-time manifold with a family of spacelike hypersurfaces 
$\Sigma_{t}$, one for each instant in time. It has been shown by the authors of Ref.~\cite{Deruelle:2009zk}
that one can decompose a gravitational action, using the ADM formalism and without necessarily moving into the Hamiltonian
regime, such that we obtain the total derivative of the gravitational action. Similar prescription allows one to seek the generalized Gibbons-Hawking-York (GHY) boundary term for IDG including Ricci scalar, tensor and Weyl, see~\cite{Teimouri:2016ulk}. 

The aim of this paper is to perform Hamiltonian analysis and identify the number of physical degrees of freedom for IDG.  
In this paper we will restrict ourselves to part of an action 
which contains only the Ricci scalar, given by  Biswas, Mazumdar and Siegel (BMS)~\cite{Biswas:2005qr}~\footnote{We will abuse the language
and call BMS action as an IDG action. However a true IDG action must have Ricci curvature and Weyl term as well, see~\cite{Biswas:2011ar,Biswas:2013kla}, and \cite{Biswas:2016etb,Biswas:2016egy}.}.
One has to determine the {\it first-class} and {\it second-class constraints}. If the Poisson bracket of a constraint with all other constraints, including itself, is equal to zero, then 
it is a {\it first-class}; otherwise, it is a {\it second-class}. We will provide examples of IDG with bad and good scenarios,  and set the criteria when an IDG can be recast in terms of finite degrees of freedom. 
As a preview, we will first consider a simple scalar field toy model with infinite derivatives, and then we will move to gravity. In the case of scalar field, one advantage will be that we will not be required to foliate the spacetime, we can recognise time direction globally.  However, when we begin our discussion of  gravity, recognising the time direction becomes vital in order to write down the conjugate momentum variables with
respect to ADM decomposition~\cite{Arnowitt:1962hi}. 

First, we will review the preliminaries of Hamiltonian analysis, provide the definitions of {\it primary, secondary, first-class} and {\it second-class constraints} and write down the formula for counting the number of degrees of freedom. In section~\ref{inscalar}, we will find the Hamiltonian and the number of degrees of freedom for scalar toy model. In section~\ref{sec:3+1}, we will describe the basics of ADM decomposition.
In section~\ref{sec:inf}, we  will illustrate the $3+1$ decomposition of an IDG theory. In section~\ref{sec:ham}, we will count the number of physical degrees of freedom for various gravitational theories, including {\it ghost free} and {\it singularity free} IDG. In section~\ref{sec:conc}, we will conclude by summarizing our results.

%%%%%%%%%%%%%%%%%%%%%%%%%%%%%%%%%%%%%%%%%%%%%%%%%%%%%%%%

\section{Hamiltonian from a Lagrangian}
\numberwithin{equation}{section}

To set the preliminaries of this paper, let us take an elementary path, suppose we have an action that depends on time evolution. We can write down the equations of motion by imposing the stationary conditions on the action and then use variational method. We start off by the following action, 
\begin{equation}
I=\int \mathcal{L} (q,\dot{q})dt\,,
\end{equation}
the above action is expressed as a time integral and $\mathcal{L}$ is the Lagrangian density depending on the position $q$ and the velocity $\dot{q}$. The variation of the action leads to  the equations of motion known as Euler-Lagrange equation,
\begin{equation}
\frac{d}{dt}\bigg(\frac{\partial\mathcal{L}}{\partial\dot{q}}\bigg)-\frac{\partial\mathcal{L}}{\partial q}=0\,,
\end{equation}
we can expand the above expression, and write, 
\begin{equation}
\ddot{q}\frac{\partial^{2}\mathcal{L}}{\partial \dot{q}\partial \dot{q}}=\frac{\partial\mathcal{L}}{\partial
q}-\dot{q}\frac{\partial^{2}\mathcal{L}}{\partial q\partial \dot{q}}\,,
\end{equation}
the above equation yields an acceleration, $\ddot{q}$, which can be uniquely calculated by position and velocity at a given time,
if and only if $\frac{\partial^{2}\mathcal{L}}{\partial \dot{q}\partial \dot{q}}$ is invertible. In other words, if the determinant  of the matrix $\frac{\partial^{2}\mathcal{L}}{\partial\dot{q}\partial \dot{q}}\neq 0$, i.e. non vanishing, then the theory is called {\it non-degenerate}. If the determinant is zero on the other hand then the acceleration can \textit{not} be uniquely determined by position and the velocity. The latter system is called {\it singular} and leads to {\it constraints} in the phase space, see~\cite{Dirac1,Wipf:1993xg,Henneaux}.

%%%%%%%%%%%%%%%%%%%%%%%%%%%%%%%%%%%%%%%%%%%%%%%%%%%%%%%%%%%%%%

\subsection{Constraints for a {\it singular} system}  

In order to formulate the Hamiltonian we need to first define the canonical momenta, 
\begin{equation}
p=\frac{\partial\mathcal{L}}{\partial\dot{q}}\,.
\end{equation}
The non-invertible matrix $\frac{\partial^{2}\mathcal{L}}{\partial
\dot{q}\partial \dot{q}}$ indicates that not all the velocities can be written in terms of the canonical momenta, in other words, not all the momenta are independent, and there are some relation between the canonical coordinates, \cite{Anderson:1951ta,Dirac:1958sq,Dirac:1958sc,Dirac1,Wipf:1993xg}, such as
\begin{equation}
\varphi(q,~p)=0\, \Longleftrightarrow {\it primary~constraints}\,,
\end{equation}
known as \textit{primary constraints}. As an example of $\varphi(q,~p)$, for instance, if we have {\it vanishing canonical momenta}, then we have {\it primary constraints}. The \textit{primary constraints} hold without using the equations of motion. The {\it primary constraints} define a submanifold smoothly embedded in a phase space, which is also known as the \textit{primary constraint surface}, 
$\Ga_p$. We can now define the Hamiltonian density as, 
\begin{equation}
\mathcal{H}=p\dot{q}-\mathcal{L}\,.
\end{equation}
If the theory admits {\it primary constraints}, we will have to redefine the Hamiltonian density,  and write the \textit{total} Hamiltonian density as, see Ref.~\cite{Dirac1}, 
\begin{equation}\label{totalH}
\mathcal{H}_{tot}=\mathcal{H}+\lambda^{a}(q,p)\varphi_{a} (q,p)\,,
\end{equation}
\\\\
where now $\lambda^{a}(q,p)$ is called the {\it Lagrange multiplier}, and $\varphi_{a} (q,p)$ are linear combinations of the primary constraints~\footnote{We should point out that the total Hamiltonian density is the sum of the canonical Hamiltonian density and terms which are products of Lagrange multipliers and the primary constraints. The time evolution of the primary constraints, should it be equal to zero, gives  the {\it secondary constraints} and those {\it secondary constraints} are evaluated by computing the Poisson bracket of the {\it primary constraints} and the total Hamiltonian density. In the literature, one may also come across the \textit{extended} Hamiltonian density, which is the sum of the canonical Hamiltonian density and terms which are products of Lagrange multipliers and the first-class constraints, see~\cite{Henneaux}.}.
The Hamiltonian equations of motion are the time evolutions, in which the Hamiltonian density remains invariant under arbitrary variations of $\delta p$, $\delta q$ and $\delta \lambda$ ; 
\begin{eqnarray}
\dot{p}=-\frac{\delta \mathcal{H}_{tot}}{\delta q}=\{q,\cH_{tot}\}\,,\\
\dot{q}=-\frac{\delta \mathcal{H}_{tot}}{\delta p}=\{p,\cH_{tot}\}\,.
\end{eqnarray}
As a result, the Hamiltonian equations of motion can be expressed in terms of the Poisson bracket.
In general, for canonical coordinates, $(q^i,~p_i)$,
on the phase space, given two functions $f(q,~p)$ and $g(q,~p)$, the Poisson
bracket can be defined as 
\begin{equation}\label{poisdef}
\{f,g\}=\sum^{n}_{i=1}\Big(\frac{\partial f}{\partial q^{i}}\frac{\partial
g}{\partial p_i}-\frac{\partial f}{\partial p_i}\frac{\partial
g}{\partial q^{i}}\Big)\,,
\end{equation} 
where $q_i$ are the generalised coordinates, and $p_i$ are the generalised conjugate momentum, and $f$ and $g$ are any function of phase space coordinates. Moreover, $i$ indicates the number of the phase space variables.

Now, any quantity is \textit{weakly} vanishing when it is numerically restricted to be zero on a submanifold $\Ga$ of the phase space, but does not vanish throughout the phase space. In other words, a function $F(p,q)$ defined in the neighbourhood of $\Ga$ is called \textit{weakly zero}, if
\be
F(p,q) |_{\Ga}=0\Longleftrightarrow F(p,q) \approx 0 \,,
\ee
where $\Ga$ is the \textit{constraint surface} defined on a submanifold of the phase space. Note that the notation ``$\approx$'' indicates that the quantity is \textit{weakly} vanishing; this is a standard Dirac's terminology, where $F(p,q)$ shall vanish on the constraint surface, $\Ga$, but not necessarily throughout the phase space.

When a theory admits \textit{primary constraints}, we must ensure that the theory is consistent by essentially 
checking whether the primary constraints are preserved under time evolution or not. In other words, we demand that, on the constraint surface $\Ga_p$, \begin{equation}\label{bvbv}
\dot\varphi |_{\Ga_p}=\{\varphi,\mathcal{H}_{tot}\}|_{\Ga_p}=0\,~~~\Longleftrightarrow \dot\varphi =\{\varphi,\mathcal{H}_{tot}\}\approx 0 \,.
\end{equation}
That is,
\be
\dot\varphi =\{\varphi,\mathcal{H}_{tot}\}\approx 0\,~~~\Longrightarrow {\it secondary~constraint} \,.
\ee
By \textit{demanding} that Eq.~\eqref{bvbv} (not identically) be zero on the constraint surface $\Ga_p$ yields a \textit{secondary constraint}~\cite{Anderson:1951ta,Matschull:1996up}, and the theory is consistent. In case, whenever Eq.~\eqref{bvbv} fixes a Lagrange multiplier, then there will be no \textit{secondary constraints}. The \textit{secondary constraints} hold when the equations of motion are satisfied, but need not hold if they are not satisfied. However, if Eq.~\eqref{bvbv} is identically zero, then there will be no {\it secondary constraints}. All constraints ({\it primary and secondary}) define a smooth submanifold of the phase space called the 
\textit{constraint surface}: $\Ga_1 \subseteq \Ga_p$.  
A theory can also admit \textit{tertiary constraints}, and so on and so forth. By satisfying the time evolution,  the procedure of finding constraints terminates after a finite number of iterations.

Note that $\mathcal{H}_{tot}$ is the total Hamiltonian density defined by Eq.~(\ref{totalH}). 
To summarize, if a canonical momentum is vanishing, we have a \textit{primary constraint}, while \textit{enforcing} that the time evolution of the {\it primary constraint}  vanishes on the constraint surface, $\Ga_1$ give rise to a \textit{secondary constraint}.

%%%%%%%%%%%%%%%%%%%%%%%%%%%%%%%%%%%%%%%%%%%%%%%%%%%%

\subsection{{\it First and second-class} constraints}

Any theory that can be formulated in Hamiltonian formalism gives rise to Hamiltonian constraints. Constraints in the context of Hamiltonian formulation can be thought of as reparameterization; while the invariance is preserved \footnote{For example, in the case of gravity, constraints are obtained by using the ADM formalism that is reparameterizing the theory under spatial and time coordinates. Hamiltonian constraints generate time diffeomorphism, see~\cite{sudarshan}.}. The most important step in Hamiltonian analysis is the classification of the constrains. By definition, we call a function $f(p,q)$ to be
{\it  first-class} if its Poisson brackets with all other constraints vanish {\it weakly}. A function which is not {\it first-class} is called
{\it second-class}~\footnote{One should mention that the {\it primary/secondary} and {\it first-class/second-class} classifications overlap. A {\it primary constraint} can be {\it first-class or second-class} and a {\it secondary constraint} can also be {\it first-class} or {\it second-class}.}. On the constraint surface $\Ga_1$, this is mathematically expressed as
\begin{eqnarray}
\left. \{f(p,q),\varphi\}\right|_{\Ga_1}\approx0 ~~\Longrightarrow~~ \textit{first-class}\,,\\
\left. \{f(p,q),\varphi\}\right|_{\Ga_1}\not\approx0 ~~\Longrightarrow~~ \textit{second-class}\,.
\end{eqnarray}
We should point out that we use the ``$\approx$'' sign as we are interested in whether the Poisson brackets of $f(p,q)$ with all other constraints vanish on the constraint surface $\Ga_1$ or not. Determining whether they vanish globally, \textit{i.e.}, throughout the phase space, is not necessary for our purposes.

%%%%%%%%%%%%%%%%%%%%%%%%%%%%%%%%%%%%%%%%%

\subsection{Counting the degrees of freedom}

Once we have the physical canonical variables, and we have fixed the number of {\it first-class and/or second-class} constraints, we can use the following formula to count the number of the physical degrees of freedom, see~\cite{Henneaux}, 
\begin{equation}\label{dofcount}
\boxed{\mathcal{N}=\frac{1}{2}(2\mathcal{A}-\mathcal{B}-2\mathcal{C})}
\end{equation}
 where
 \begin{itemize}
 \item{$\mathcal{N}=$ number of physical degrees of freedom}
 \item{ $\mathcal{A}=$  number of \textit{{phase space variables}}} 
 \item{$\mathcal{B}=$  number of \textit{{second-class constraints}}} 
 \item{$\mathcal{C}=$  number
of \textit{{first-class constraints}}}
\end{itemize}
 
%%%%%%%%%%%%%%%%%%%%%%%%%%%%%%%%%%%%%%%%%%%%%%%%%%%%%%%%%%

\section{Infinite derivative scalar field theory}\label{inscalar}

In this section and before moving on to gravity we are going to consider a  Lagrangian which is constructed by infinite number of d'Alembertian operators, 
in other words, we can have an action of the form, in Minkowski spacetime, 
\begin{equation}\label{inftscalar}
I_{}=\int d^{4}x \, \phi\mathcal{F}(\bbox)\phi, \qquad\text{with:}\qquad\mathcal{F}(\bbox)=\sum^{\
\infty}_{n=0}c_{n}\bbox^n\,,
\end{equation}
where $c_n$ are constants. Of course, for the above action, we would need a slightly more sophisticated approach, see the analysis by \cite{Kluson:2011tb}, which we follow here  by first writing an equivalent action of the form, 
\begin{eqnarray}\label{eqvaction1}
I_{eqv}=\int d^{4}x A\mathcal{F}(\bbox)A\,,\qquad\text{with:}\qquad \mathcal{F}(\bbox)A=\sum^{\
\infty}_{n=0}c_{n}\bbox^{n}A
\end{eqnarray} 
Where the auxiliary field, $A$, is introduced as an equivalent scalar field to  $\phi$, this means that the equations of the 
motion for both actions ($I$ and $I_{eqv}$) are equivalent. 
 
Now, in order to eliminate the contribution of $\bbox A,~\bbox^{2}A$ and so
on, we are going to introduce two auxiliary fields $\chi_n$ and $\eta_n$,  where the $\chi_{n}$'s are dimensionless and the $\eta_{n}$'s have mass dimension $2$ (this can be seen by parameterising $\bbox A$, $\bbox^{2} A$, $\cdots$). We show few steps here by taking a simple example

\begin{itemize}
\item
Suppose our action is built by one box only, then, 
\begin{equation}\label{blabla1}
I_{eqv}=\int d^{4}x A\bbox A \,.
\end{equation}
Now,  to eliminate $\bbox A$ in the term $A\bbox A$, we wish to add a following term in the
above action, 
\begin{equation}\label{additionalterm}
\int d^{4}x~\chi_1A(\eta_1-\bbox A)=\int d^{4}x\bigg[\chi_1
A\eta_1+g^{\mu\nu}(\partial_\mu\chi_1 A \partial_\nu A+\chi_1
\partial_\mu A
\partial_\nu A)\bigg]\,.
\end{equation}   
and hence, 
we have, 
\begin{equation}\label{blabla2}
I_{eqv}=\int d^{4}x\Big( A\eta_{1}+\chi_1A(\eta_1-\bbox A)\Big)\,,
\end{equation}
by solving the equation of motion for $\chi_1$, we obtain 
\begin{equation}
\eta_1=\bbox A\,,
\end{equation} and hence, Eqs. (\ref{blabla1}) and (\ref{blabla2}) are equivalent.
\end{itemize}

\noindent
Similarly, in order to eliminate the terms $A\bbox^{n}A$ and so on,
we have to repeat the same procedure up to $\bbox^n$. Note that we have established this by
solving the equation of motion for $\chi_{n}$, we obtain, for $n \geq 2$, 
\begin{equation}
\eta_{n}=\bbox \eta_{n-1}=\bbox^{n}A.
\end{equation}
Now, we can rewrite the action Eq.~(\ref{eqvaction1}) as,
\begin{eqnarray}\label{eqvv2}
I_{eqv}&=&\int d^{4}x\Bigg\{A(c_{0}A+\sum^{\infty}_{n=1}c_n\eta_{n})+\chi_{1}A(\eta_{1}-\Box
A) +\sum
^{\infty}_{l=2}\chi_lA(\eta_l-\Box \eta_{l-1})\Bigg\}\nonumber\\
&=&\int d^{4}x\Bigg\{A(c_{0}A+\sum^{\infty}_{n=1}c_n\eta_{n})+\sum
^{\infty}_{l=1}A\chi_l
\eta_l\nonumber\\
&+&\eta^{00}(A\partial_0\chi_1  \partial_0 A+\chi_1
\partial_0 A
\partial_0 A)+\eta^{ij}(A\partial_i\chi_1  \partial_j A+\chi_1
\partial_i A
\partial_j A)\nonumber\\
&+&\eta^{00}\sum^{\infty}_{l=2}(A\partial_0\chi_{l}  \partial_0\eta_{l-1}
+\chi_l
\partial_0 A
\partial_0 \eta_{l-1})+\eta^{ij}\sum^{\infty}_{l=2}(A\partial_i\chi_{l}  \partial_j\eta_{l-1}
+\chi_l
\partial_i A
\partial_j \eta_{l-1}) 
\Bigg\}\,.\nonumber\\
\end{eqnarray}
where we have absorbed the powers of $M^{-2}$ into the $c_{n}$'s \& $\chi_{n}$'s and the mass dimension of the $\eta_{n}$'s has been modified accordingly.
Hence, the
box operator is not barred. 
We also  decomposed the d'Alembertian operator to its components around the Minkowski background: $\Box=\eta^{\mu\nu}\partial_\mu\partial_\nu=\eta^{00}\partial_0\partial_0+\eta^{ij}\partial _{i}\partial_{j}$, where the zeroth component is the time coordinate, and $\{i,~j\}$ are the spatial coordinates running from $1$ to $3$. 
The conjugate momenta for the above action are given by:
\begin{eqnarray}
&&p_A=\frac{\partial\mathcal{L}}{\partial\dot{A}}=\Big[-(A\partial_0\chi_1
 +\chi_1
\partial_0 A)-\sum^{\infty}_{l=2}(\chi_l\partial_0 \eta_{l-1})\Big],\quad
\nonumber\\
&&p_{\chi_1}=\frac{\partial\mathcal{L}}{\partial\dot{\chi}_1}=-A\partial_0
A,\quad p_{\chi_l}=\frac{\partial\mathcal{L}}{\partial\dot{\chi}_l}=-(A\partial_0\eta_{l-1}),\quad\nonumber\\
&&p_{\eta_{l-1}}=\frac{\partial\mathcal{L}}{\partial\dot{\eta}_{l-1}}=-(A\partial_0\chi_l
+\chi_l
\partial_0 A).
\end{eqnarray}
where $\dot{A}\equiv\partial_0 A$.  Therefore, the Hamiltonian density is given by (see Appendix \ref{hamdensapp} for explicit derivation
Eq.~(\ref{hamapp-1})):
\begin{eqnarray}\label{hamdens}
\mathcal{H}&=&p_A\dot{A}+p_{\chi_1}\dot{\chi}_1+p_{\chi_l}\dot{\chi}_l+p_{\eta_{l-1}}\dot{\eta}_{l-1}-\mathcal{L}\nonumber\\&=&
A(c_{0}A+\sum^{\infty}_{n=1}c_n\eta_{n})-\sum ^{\infty}_{l=1}A\chi_l
\eta_l\nonumber\\
&-&(\eta^{\mu\nu}A\partial_{\mu}\chi_1  \partial_\nu A+\eta^{ij}\chi_1
\partial_i A
\partial_j A)
-\eta^{\mu\nu}\sum^{\infty}_{l=2}(A\partial_\mu\chi_{l}  \partial_\nu\eta_{l-1}
+\chi_l
\partial_\mu A
\partial_\nu \eta_{l-1})\,.\nonumber\\
\end{eqnarray}
Let us now consider the first line of Eq.~(\ref{eqvv2}), before integrating by parts. We see that we
have  terms like :  $$\chi_{1}A(\eta_{1}-\Box A)$$ and $$\chi_lA(\eta_l-\Box
\eta_{l-1}),~~for~~l \geq 2.$$ Moreover,
we know that solving the equations of motion for $\chi_n$ leads to $\eta_n=\Box^{n}
A$. Therefore, we can conclude that the $\chi_{n}$'s are the Lagrange multipliers, they do not appear in the dynamics, 
and therefore from the equations of motion, we get the following {\it primary constraints}~\footnote{ Let us note that $\Ga_p$ is a smooth submanifold of the phase space determined by the {\it primary constraints}; in this section, we shall exclusively use the ``$\approx$'' notation to denote equality on $\Ga_p$.}:
\begin{align}
\sa_{1} =\eta_{1}-\Box A & \approx 0 \,, \nonumber \\
&\vdots \\ \nonumber
\sa_{l}=\eta_{l}-\Box \eta_{l-1} & \approx 0 \,.
\end{align} 
In other words, since $\chi_{n}$'s are the Lagrange multipliers, therefore $\sa_{1}$ and $\sa_{l}$'s are {\it primary constraints}.
The time evolutions of the $\sa_n$'s fix the corresponding Lagrange multipliers $\la^{\sa_n}$ in the total Hamiltonian (when we add the terms $\la^{\sa_n}\sa_{n}$
to the Hamiltonian density $\cH$); therefore, the $\sa_n$'s do not induce \textit{secondary constraints}.
As a result, to classify the above constraint, we will need to show that the Poisson bracket given in Eq. (\ref{poisdef}) {\it weakly} vanishes:
\begin{equation}
\{\sigma _{m},\sigma _{n}\}|_{\Ga_p}= 0\,,
\end{equation}
such that $\sa_{n}$'s can be classified as {\it first-class constraints}. However, this depends on the choice of ${\cal F}(\Box)$, whose coefficients are hiding in $\chi$'s and $\eta$'s.
It is trivial to show that, for this case, there is no {\it second-class constraint}, \textit{i.e.}, ${\cal B}=0$, as we do not have $\{\sigma _{m},\sigma _{n}\}\not\approx0$. That is, the $\sa_n$'s are \textit{primary, first-class constraints}.  
In our case, the number of  phase space variables, 
\begin{eqnarray}\label{phasespacevariables2}
2\mathcal{A}&\equiv&2\times\bigg\{(A,p_A),\underbrace{(\eta_1,p_{\eta_1}),(\eta_2,p_{\eta_2}),\cdots}_{n=1,~2,~3,\cdots \infty}\bigg\}\equiv2\times(1+\infty)=2+\infty\,. \nonumber \\
\end{eqnarray} 
For each pair,  $(\eta_n,~p_{\eta_n})$, 
we have assigned one variable, which is multiplied by a factor of $2$, since we
are dealing with field-conjugate momentum pairs, in the phase space. In the next section, we will fix the form of ${\cal F}(\bar\Box)$
to estimate the number of {\it first-class constraints}, \textit{i.e.}, ${\cal C}$ and, hence, the number of degrees of freedom. Let us also mention that the choice of $\cF(\bbox)$ will determine the number of solutions to the equation of motion for $A$ we will have, and consequently these solutions can be interpreted as {\it first-class constraints} which will determine the number of physical degrees of freedom., \textit{i.e.}  finite/infinite number of degrees of freedom will depend on the number of solutions of the equations of motion for $A$.

%%%%%%%%%%%%%%%%%%%%%%%%%%%%%%%%%%%%%%%%%%%%%%%%%%%%%

\subsection{Gaussian kinetic term and propagator}

Let us now consider an example of infinite derivative scalar field theory, but with a Gaussian kinetic term in Eq. (\ref{inftscalar}), \textit{i.e.}, 
by exponential of an entire function,  
\begin{equation}
\label{action-entire}
I_{eqv}=\int d^{4} x~A\bigg(\Box e^{-\bbox}\bigg)A\,.
\end{equation}
For the above action, the equation of motion for $A$ is then given by:
\begin{equation}
2\bigg(\Box e^{-\bbox}\bigg)A=0\,.
\end{equation}
We observe that there is a finite number of solutions; hence, there are also finitely many degrees of freedom~\footnote{Note that, for an infinite derivative action of the form $I_{eqv}=\int d^{4} x~A\cos (\bbox) A$, we would have an infinite number of solutions and, hence, infinitely many degrees of freedom.}. In momentum space, we obtain the following solution, 
\begin{equation}
k^{2}=0\,,
\end{equation}
and the propagator will follow as~\cite{Biswas:2011ar,Biswas:2013kla} :
\begin{equation}
\Pi(\kb^2)\sim \frac{1}{k^{2}}e^{-\kb^2}\,,
\end{equation}
where we have used the fact that in momentum space $\Box\rightarrow -k^2$, and
we have $\bar{k}\equiv k/M$. There are some interesting properties to note about this propagator  

\begin{itemize}

\item{The propagator is suppressed by an {\it exponential of an entire function}, which has no zeros, or no poles. 
Therefore, the only dynamical pole resides at $k^{2}=0$, \textit{i.e.}, the massless pole in the propagator, \textit{i.e.}, degrees of freedom ${\cal A} =1$. In spite of having infinitely many derivatives, the theory has maintained that the only $1$ relevant degrees of freedom is the massless scalar field. In fact, there are 
no new dynamical degrees of freedom. Furthermore, in the UV the propagator is suppressed~\cite{Talaganis:2014ida}. }

\item{The propagator contains no {\it ghosts}, which usually plagues higher derivative theories. By virtue of this, at a classical 
level there is no analogue of Ostr\'ogradsky instability. We will discuss briefly in an appendix how to solve the infinite derivative 
equation of motion. Given the background equation, one can indeed understand the stability of the solution. Such studies have been 
performed in past in connection for IDG in the context of cosmology, see~\cite{Biswas:2010zk,Biswas:2012bp}.}

\end{itemize}

The original action Eq.~(\ref{action-entire}) can now be recast in terms of an equivalent action, very similar to what we have discussed 
in the previous subsection, as:
\begin{eqnarray}
I_{eqv}&=&\int d^{4} x\bigg[A\big(\Box e^{-\bbox}\big)A+\chi_{1}A(\eta_{1}-\Box A) +\sum
^{\infty}_{l=2}\chi_lA(\eta_l-\Box \eta_{l-1})\bigg]\,.\nonumber\\
\end{eqnarray}
We can  now compute the number of the physical degrees of freedom. Note that the determinant of the phase-space dependent matrix 
$A_{mn}=\{\sa_{m},\sa_{n} \}\neq 0$, so the $\sa_n$'s do not induce further constraints, such as {\it secondary constraints}. Therefore,
\begin{eqnarray}\label{dog}
&&2\mathcal{A}\equiv2\times\bigg\{(A,p_A),\underbrace{(\eta_1,p_{\eta_1}),(\eta_2,p_{\eta_2}),\cdots}_{n}\bigg\}
=2\times(1+\infty)=2+\infty\nonumber\\
&&\mathcal{B}=0,\nonumber\\
&&2\mathcal{C}\equiv2\times(\sigma
_{n})=2(\infty)=\infty,\nonumber\\
&&\mathcal{N}=\frac{1}{2}(2\mathcal{A}-\mathcal{B}-2\mathcal{C})=\frac{1}{2}(2+\infty-0-\infty)=1\,.
\end{eqnarray}
As expected, the conclusion of this analysis yields exactly the same dynamical degrees of freedom as that of the Lagrangian formulation. 
The coefficients  $c_i$ of ${\cal F}(\bbox)$ are all fixed by the form of $\Box e^{-\bbox}$.

%%%%%%%%%%%%%%%%%%%%%%%%%%%%%%%%%%%%%%%%%%%%%%%%%%%%%%%%%

\section{Infinite derivative gravity (IDG)}
\numberwithin{equation}{section}
\label{sec:3+1}

In this section we will take a simple action of IDG~\cite{Biswas:2005qr}, and study the Hamiltonian density and degrees of freedom, but before that we 
briefly recap the ADM formalism for gravity as we will require this for further development.

\subsection{ADM formalism}
One of the important concepts in GR is diffeomorphism invariance, i.e. when one transforms coordinates at given space-time points, the physics remains unchanged. As a result of this, one concludes that diffeomorphism is a local transformation.   
In Hamiltonian formalism, we have to specify the direction of time. A very useful approach to do this is ADM decomposition~\cite{Arnowitt:1962hi,Gourgoulhon:2007ue}, such decomposition permits to choose one specific time direction without violating the diffeomorphism invariance. In other words, choosing the time direction is nothing but gauge redundancy, or making sure that diffeomorphism is a local transformation. We assume  that the manifold $\mathcal{M}$ is a time orientable spacetime, which
can be foliated by a family of space like hypersurfaces $\Sigma_{t}$, at
which the time is fixed to be constant $t=x^{0}$. We then introduce an induced
metric on the hypersurface as $$h_{ij}\equiv g_{ij}|_{_{t}}\,,$$ where the Latin
indices run from 1 to 3 for spatial coordinates. 

In $3+1$ formalism
the line element is parameterised as, 
\begin{eqnarray}\label{metric1}
ds^{2}=-N^{2}dt^{2}+h_{ij}(dx^i+N^i dt)(dx^{j}+N^j dt)\,,
\end{eqnarray}
where $N$ is the \textit{lapse function} and $N^{i}$ is the \textit{shift vector},
given by
\begin{equation}\label{lapseshift}
N=\frac{1}{\sqrt{-g^{00}}},\qquad N^i=-\frac{g^{0i}}{g^{00}}\,.
\end{equation} 
In terms of metric variables, we then have
\begin{eqnarray}
&&g_{00}=-N^2+h_{ij}N^iN^j,\qquad g_{0i}=N_i, \qquad g_{ij}=h_{ij}\,,\nonumber\\
&&g^{00}=-N^{-2},\qquad g^{0i}=\frac{N^i}{N^2}, \qquad g^{ij}=h^{ij}-\frac{N^{i}N^{j}}{N^{2}}\,.
\end{eqnarray} 
Furthermore, we can define $n^\mu$ to be the vector normal to the hypersurface. For the time like vector $n^{\mu}$ in Eq. (\ref{metric1}), they take the following
form:
\begin{eqnarray}\label{normvar}
n_{i}=0, \quad n^{i}=-\frac{N^{i}}{N},\quad
n_{0}=-N,\quad\ n^0=N^{-1}\,.
\end{eqnarray}
From Eq.~(\ref{metric1}), we also get $\sqrt{-g}=N\sqrt{h}$.

In addition, we are going to introduce a covariant derivative associated with the induced
metric $h_{ij}$:
$$D_i\equiv e^{\mu}_{i}\nabla_\mu\,.$$
  We will  define the extrinsic curvature as:
\begin{equation}\label{extrinsiccurvature}
K_{ij}=-\frac{1}{2N}\left(D_{i}N_{j}+D_{j}N_{i}-\partial_{t}h_{ij}\right)\,.
\end{equation} 
It is well known that the Riemannian curvatures can be written in terms of
the 3+1 variables. In the case of scalar curvature we have~\cite{Gourgoulhon:2007ue}: 
\begin{eqnarray}\label{scalarcurv}
R=K_{ij}K^{ij}-K^2+\mathcal{R}+\frac{2}{\sqrt{h}}\partial_\mu(\sqrt{h}n^\mu
K)-\frac{2}{N\sqrt{h}}\partial_i(\sqrt{h}h^{ij}\partial_j N)\,,
\end{eqnarray} 
where $K=h^{ij}K_{ij}$ is the trace of the extrinsic curvature, and $\mathcal{R}$
is scalar curvature calculated using the induced metric $h_{ij}$~\footnote{We note
that the Greek indices are $4$-dimensional while Latin indices are spatial
and $3$-dimensional.}. 

One can calculate each term in Eq.~(\ref{scalarcurv})
using the information about extrinsic curvature and those provided in Eq.~(\ref{normvar}).
The decomposition of the d'Alembertian operator can be expressed as:
\begin{eqnarray}\label{boxdecomposition}
\Box&=&g^{\mu\nu}\nabla_\mu\nabla_\nu\\\nonumber
&=&(h^{\mu\nu}+\varepsilon n^\mu n^\nu)\nabla_\mu\nabla_\nu=(h^{ij}e^{\mu}_{i}e^{\nu}_{j}-
n^\mu n^\nu)\nabla_\mu\nabla_\nu\\\nonumber
&=&h^{ij}D_iD_j-n^\nu\nabla_\bn\nabla_\nu=\Box_{hyp}-n^\nu\nabla_\bn\nabla_\nu\,,
\end{eqnarray} 
where we have used the completeness relation for a spacelike hypersurface, i.e.
$\varepsilon=-1$, and we have defined $\nabla_\bn=n^\mu\nabla_\mu$.

%%%%%%%%%%%%%%%%%%%%%%%%%%%%%%%%%%%%%%%%%%%%%%%%

\subsection{ADM decomposition of an Infinite derivative gravity}
\numberwithin{equation}{section}
\label{sec:inf}

Let us now introduce an action for an IDG. In this paper we will restrict ourselves to part of an action 
which contains on the Ricci scalar, or the BMS action~\cite{Biswas:2005qr}:
\begin{eqnarray}\label{eq1}
S=\frac{1}{2}\int d^{4}x \, \sqrt{-g}\bigg[M^{2}_{p}R+R\mathcal{F}(\bbox)R\bigg],\qquad \mathcal{F}(\bbox)=\sum^{\
\infty}_{n=0}f_{n}\bbox^{n}\,,
\end{eqnarray}
where $M_{p}$ is the $4$-dimensional Planck scale, given by $M^{2}_{p}=(8\pi G)^{-1}$, with $G$ is Newton's gravitational constant. The first term is Einstein Hilbert term, with $R$ being scalar curvature in four dimensions and the second term is the infinite derivative modification to the action, where $\bbox\equiv\Box/M^{2}$ ,  since $\Box$ has dimension mass squared and $\mathcal{F}(\bbox)$ will be dimensionless. Note that $\Box$ is the $4$-dimensional d'Alembertian operator given by $\Box=g^{\mu\nu}\nabla_{\mu}\nabla_{\nu}$. Moreover, $f_{n}$ are the dimensionless coefficients of the series expansion.

%%%%%%%%%%%%%%%%%%%%%%%%%%%%%%%%%%%%%%%%%%%%%%%%%%%%%%%%

\subsection{Equivalent action and decomposition}\label{idgsec}

Now that the pillars of the $3+1$ decomposition are set, we  rewrite our original action given in Eq.(\ref{eq1})
in its equivalent form. We start off by writing an equivalent action as, see~\cite{Biswas:2005qr}
\begin{eqnarray}\label{eqvaction}
S_{eqv}=\frac{1}{2}\int d^{4}x \, \sqrt{-g}\bigg[M^{2}_{p}A+A\mathcal{F}(\bbox)A+B(R-A)\bigg]\,,
\end{eqnarray}
where we have introduced two scalar fields $A$ and $B$ with mass dimension two. Solving the equations of motion for scalar field $B$ results in $A=R$. The equations of motion for the original action, Eq.(\ref{eq1}),  are equivalent to the equations of motion for Eq. (\ref{eqvaction}).
\begin{eqnarray}
\delta S_{eqv}=\frac{1}{2}\delta\Bigg\{\sqrt{-g}\bigg[M^{2}_{p}A+A\mathcal{F}(\bbox)A+B(R-A)\bigg]\Bigg\}=0\Rightarrow R=A\,.
\end{eqnarray}
Following the steps of a scalar field theory, we  expand $\mathcal{F}(\bbox)A$, using Eq. (\ref{eq1}):
\begin{equation}
\mathcal{F}(\bbox)A=\sum^{\
\infty}_{n=0}f_{n}\bbox^{n}A=f_{0}A+f_{1}\bbox A+f_{2}\bbox^{2}A+f_{3}\bbox^{3}A+\cdots
\end{equation} 
As before, in order to eliminate  $\bbox A,~\bbox^{2}A, \cdots$, we will introduce two new auxiliary fields $\chi_n$ and $\eta_n$ with 
the $\chi_{n}$'s being dimensionless and the $\eta_{n}$'s of mass dimension two~\footnote{This part of the discussion is very similar to case of infinite derivative scalar field theory, see section \ref{inscalar}.}. 
Thus, we can rewrite the action Eq.~(\ref{eqvaction}), as:
\begin{align}\label{eqv2}
S_{eqv}&=\frac{1}{2}\int d^{4}x\sqrt{-g}\Bigg\{A(M^{2}_{p}+f_{0}A+\sum^{\infty}_{n=1}f_n\eta_{n})+B(R-A)+\chi_{1}A(\eta_{1}-\Box A)\non &
+\sum
^{\infty}_{l=2}\chi_lA(\eta_l-\Box \eta_{l-1})\Bigg\}
\nonumber\\&=
\frac{1}{2}\int d^{4}x\sqrt{-g}\Bigg\{A(M^{2}_{p}+f_{0}A+\sum^{\infty}_{n=1}f_n\eta_{n})+B(^{}R-A)\nonumber\\
&+g^{\mu\nu}(A\partial_\mu\chi_1  \partial_\nu A+\chi_1
\partial_\mu A
\partial_\nu A)+g^{\mu\nu}\sum^{\infty}_{l=2}(A\partial_\mu\chi_l  \partial_\nu\eta_{l-1}
+\chi_l
\partial_\mu A
\partial_\nu \eta_{l-1}) \nonumber \\
&+\sum ^{\infty}_{l=1}A\chi_l \eta_l 
\Bigg\}\,,
\end{align}
where we have absorbed the powers of $M^{-2}$ into the $f_{n}$'s and $\chi_{n}$'s,
 and the mass dimension of $\eta_{n}$'s has been modified accordingly, hence, the box operator is not barred.

Note that the gravitational part of the action is simplified. 
In order to perform the ADM decomposition, let us first look at the $B(R-A)$ term, with the help of Eq. (\ref{scalarcurv})
we can write:
\begin{eqnarray}\label{bra}
B(R-A)=B\Big(K_{ij}K^{ij}-K^2+\mathcal{R}-A\Big)-2\nabla_\bn BK-\frac{2}{\sqrt{h}}\partial_j(\partial_i(B)\sqrt{h}h^{ij}
)\,,\nonumber \\
\end{eqnarray}
where we have used $n^\mu\nabla_\mu\equiv\nabla_\bn$ and dropped the total derivatives (See Appendix \ref{app2} for relevant steps). 
Furthermore, we can use the decomposition of d'Alembertian operator, given in Eq. (\ref{boxdecomposition}), and also in 3+1, we have 
$\sqrt{-g}=N\sqrt{h}$. Hence, the decomposition of Eq. (\ref{eqv2}) becomes:
\begin{eqnarray}\label{finaldecomposition}
S_{eqv}^{'}&=&\frac{1}{2}\int d^{3}xN\sqrt{h}\Bigg\{A(M^{2}_{p}+f_{0}A+\sum^{\infty}_{n=1}f_n\eta_{n})+B\Big(K_{ij}K^{ij}-K^2+\mathcal{R}-A\Big)\nonumber\\
&-&2\nabla_\bn BK-\frac{2}{\sqrt{h}}\partial_j(\partial_i(B)\sqrt{h}h^{ij}
)
\nonumber\\&+&h^{ij}(A\partial_i\chi_1  \partial_j A+\chi_1
\partial_i A
\partial_j A)-(A\nabla_\bn \chi_1  \nabla_\bn A+\chi_1
\nabla_\bn A\nabla_\bn A)\nonumber\\&+&h^{ij}\sum^{\infty}_{l=2}(A\partial_i\chi_l  \partial_j\eta_{l-1}
+\chi_l
\partial_i A
\partial_j \eta_{l-1})-\sum^{\infty}_{l=2}(A\nabla_\bn \chi_l  \nabla_\bn \eta_{l-1}
+\chi_l
\nabla_\bn  A\nabla_\bn \eta_{l-1})\nonumber\\&+&\sum ^{\infty}_{l=1}A\chi_l \eta_l
\Bigg\}\,,
\end{eqnarray}
where  the Latin indices are spatial, and run from 1 to 3. Note that the $\chi$ fields were introduced to parameterise the contribution of 
$\bbox A,~\bbox^2A,~\cdots $, and so on. Therefore, $A$ and $\eta$ are auxiliary fields, which concludes that $\chi$ fields have no intrinsic value, and hence to be made redundant, i.e. Lagrange multiplier, when we count the number of phase space variables. 

The same can not be concluded regarding the $B$ field, as it is introduced to obtain equivalence between scalar curvature, $R$, and $A$. Since $B$ field is coupled to $R$, and the Riemannian curvature is physical - we must count $B$ as a phase space variable. As we will see later in our Hamiltonian analysis, this is a crucial point while counting the number of physical degrees of freedom correctly. To summarize, as we will see below, $B$ field is not a Lagrange multiplier, while $\chi$ fields are.

%%%%%%%%%%%%%%%%%%%%%%%%%%%%%%%%%%%%%%%%%%%%%%%%%%%%%%%

\section{Hamiltonian analysis}
\numberwithin{equation}{section}
\label{sec:ham}

Let us warm up with some simple examples in gravity where we determine the dynamical degrees of freedom by 
using Eq.~(\ref{dofcount}), in particular $f(R)$ gravity~\footnote{ A pedagogical discussion on Hamiltonian analysis can be found in Ref.~\cite{Sotiriou:2008rp}, and 
Ref.~\cite{Deruelle:2009zk}. A similar technique have been employed to find the boundary terms for an infinite derivative gravity for 
Ricci scalar, Ricci tensor and Riemann tensor, see Ref.~\cite{Teimouri:2016ulk}.}.
 
%%%%%%%%%%%%%%%%%%%%%%%%%%%%%%%%%%

\subsection{Hamiltonian for $f(R)$ gravity}

The action of $f(R)$ gravity is given by, 
\begin{eqnarray}\label{fraction}
S=\frac{1}{2\kappa}\int d^{4}x\sqrt{-g}f(R)\,,
\end{eqnarray}
where $f(R)$ is a function of scalar curvature and $\kappa=8\pi G$. The equivalent action for above is then given by, 
\begin{eqnarray}
S=\frac{1}{2\kappa}\int d^{4}x\sqrt{-g}\Big(f(A)+B(R-A)\Big)\,,
\end{eqnarray}
where again solving the equations of motion for $B,$ one obtains $R=A$, and hence it is clear that above action is equivalent 
with Eq. (\ref{fraction}). Using Eq. (\ref{bra}) we can  decompose the action as,
\begin{eqnarray}
S^{'}=\frac{1}{2\kappa}\int d^{3}xN\sqrt{h}\Big(f(A)+B\Big(K_{ij}K^{ij}-K^2+\mathcal{R}-A\Big)-2\nabla_\bn
BK\nonumber \\
-\frac{2}{\sqrt{h}}\partial_j(\partial_i(B)\sqrt{h}h^{ij}
)\Big)\,.
\end{eqnarray}
Now that the above  action is expressed in terms of $(h_{ab},~N,~N^i,~B,~A)$, and their time and space derivatives. 
We can proceed with the Hamiltonian analysis and write down the momentum conjugate for each of these variables:
\begin{eqnarray}\label{const-1}
&&\pi^{ij}=\frac{\partial\mathcal{L}}{\partial\dot{h}_{ij}}=\sqrt{h}B(K^{ij}-h^{ij}K)-\sqrt{h}\nabla_\bn Bh^{ij},\qquad p_B=\frac{\partial\mathcal{L}}{\partial\dot{B}}=-2\sqrt{h}K,\nonumber\\
&&p_A=\frac{\partial\mathcal{L}}{\partial\dot{A}}\approx0, \qquad\pi_N=\frac{\partial\mathcal{L}}{\partial\dot{N}}\approx0,\qquad\ \pi_i=\frac{\partial\mathcal{L}}{\partial\dot{N^i}}\approx0\,.
\end{eqnarray}
where $\dot{A}\equiv\partial_0A$ is the time derivative of the variable.
We have used the ``$\approx$'' sign in Eq.~\eqref{const-1} to show that $(p_A,~\pi_N,~\pi_i)$ are \textit{primary constraints} satisfied on the constraint surface: $$\Ga_p=(p_A \approx 0,~\pi_N\approx 0,~\pi_i \approx 0),$$ defined by the aforementioned {\it primary constraints}. For our purposes, whether the {\it primary constraints} vanish globally (which they do), \textit{i.e.}, throughout the phase space, is irrelevant. Note that the Lagrangian density, $\mathcal{L}$, does not contain  $\dot A$, $\dot N$ or $\dot N^i$, therefore, 
their conjugate momenta vanish identically. 

We can define the Hamiltonian density as~\footnote{We should note that, in Ref.~\cite{Kluson:2011tb}, the notation $\cH_T$ is used for the quantity $\da \cH / \da N$ which we denote by $\cH_N$, \textit{i.e.}, the variational derivative of the canonical Hamiltonian density $\cH$ with respect to the lapse function $N$. In the literature, the notation $\cH_T$ is, sometimes, used to refer to the total Hamiltonian density; in~\cite{Kluson:2011tb}, the notation $\cH_T$ does \textit{not} refer to the total Hamiltonian density. In this paper, we use the notation $\cH_{tot}$ for the total Hamiltonian density.}: 
\begin{align}\label{hamdensity1}
\mathcal{H}&=\pi^{ij}\dot{h}_{ij}+p_B\dot{B}-\mathcal{L}\\
&\equiv\ N\mathcal{H}_N+N^i\cH_i \,,
\end{align}
where  $\mathcal{H}_N=\dot\pi_N$, and $\cH_i=\dot{\pi}_{i}$.
After some algebra  and using Eq. (\ref{hamdensity1}), we can write 
\begin{eqnarray}\label{h1}
&&\mathcal{H}_N=\frac{1}{\sqrt{h}B}\pi^{ij}h_{ik}h_{jl}\pi^{kl}-\frac{1}{3\sqrt{h}B}\pi^2-\frac{\pi
p_B}{3\sqrt{h}}+\frac{B}{6\sqrt{h}}p_B^{2}\nonumber\\&&-\sqrt{h}BR+\sqrt{h}BA+2\partial_j[\sqrt{h}h^{ij}\partial_i]B+f(A)\,,
\end{eqnarray}
 and, 
\begin{equation}\label{h2}
\cH_i=-2h_{ik}\nabla_l\pi^{kl}+p_B\partial_iB\,.
\end{equation}
Therefore, the total Hamiltonian can be written as  (in terms of Lagrange multipliers):
\begin{align}
H_{tot}&=\int d^3 x \, \cH \\
&=\int d^{3}x \, \Big(N\mathcal{H}_N+N^i\cH_i+\lambda^{A}p_A+\lambda^{N}\pi_N+\lambda^{i}\pi_i\Big)\,,
\end{align}
where  $\lambda^{A},\lambda^{N},\la^{i}$ are Lagrange multipliers, and we have  $G_A=\dot p_A$.   

%%%%%%%%%%%%%%%%%%%%%%%%%%%%%%%%%%%%%%%%%%%%%%%%

\subsubsection{Classification of constraints for $f(R)$ gravity}

Having vanishing conjugate momenta means we can not express $\dot{A}$, $\dot{N}$ and $\dot{N}^i$ as a function 
of their conjugate momenta and hence $p_A\approx0,~\pi_N\approx0$ and $\pi_i\approx0$ are \textit{primary constraints},
see Eq.(\ref{const-1}). To ensure the consistency of the {\it primary constraints} so that they are preserved under time evolution 
generated by total Hamiltonian $H_{tot}$, we need to employ the Hamiltonian
field equations and \textit{enforce} that $\cH_{N}$ and $\cH_{i}$ be zero on the constraint surface $\Ga_p$,
\begin{equation}\label{secondaryconstraints1}
\dot{\pi}_{N}=-\frac{\delta \mathcal{H}_{tot}}{\delta N}=\cH_{N}\approx0,\qquad \dot{\pi}_{i}=-\frac{\delta \mathcal{H}_{tot}}{\delta N^i}=\cH_{i}\approx0\,,
\end{equation}
such that $\mathcal{H}_{N}\approx0$ and $H_{i}\approx0$, and therefore they can be treated as \textit{secondary constraints}. 

Let us also note that $\Ga_1$ is a smooth submanifold of the phase space determined by the {\it primary} and  {\it secondary constraints}; hereafter in this section, we shall exclusively use the ``$\approx$'' notation to denote equality on $\Ga_1$. 
It is usual to call $\mathcal{H}_{N}$ as the \textit{Hamiltonian constraint}, and $\cH_i$ as  \textit{diffeomorphism constraint}. Note that  $\mathcal{H}_{N}$ and $\cH_{i}$ are weakly vanishing \textit{only} on the constraint surface; this is why the r.h.s of Eqs. (\ref{h1}) and (\ref{h2}) are not  {\it identically} zero. If $\dot{\pi}_{N}=\cH_{N}$ and $\dot{\pi}_{i}=\cH_{i}$ were \textit{identically} zero, then there would be no {\it secondary constraints}.  

Furthermore, we are going to define $G_A$, and \textit{demand} that $G_A$ be weakly zero on the constraint surface 
$\Ga_1$,
\begin{equation}
G_{A}=\partial_tp_A=\{{p_A},\mathcal{H}_{tot}\}=-\frac{\delta \mathcal{H}_{tot} }{\delta A}=-\sqrt{h}N(B+f'(A))\approx0\,,
\end{equation}
which will act as a {\it secondary constraint} corresponding to {\it primary constraint} $p_A\approx0$.  Hence, 
$$\Ga_1=(p_A \approx 0,~\pi_N \approx 0,~\pi_i \approx 0,~G_A \approx 0,~{\cal H}_{N}\approx 0,~{\cal H}_{i}\approx 0).$$

Following the definition of Poisson bracket in Eq.(\ref{poisdef}), we can see that since the constraints $\cH_N$ and $\cH_i$ are preserved under time evolution, \textit{i.e.}, $\dot \cH_N=\{\cH_N,\mathcal{H}_{tot}\}|_{\Ga_1}= 0$ and $\dot \cH_i=\{\cH_i,\mathcal{H}_{tot}\}|_{\Ga_1}= 0$, and they fix the Lagrange multipliers $\la^N$ and $\la^i$. That is, the expressions for $\dot \cH_N$ and $\dot \cH_i$ include the Lagrange multipliers $\la^N$ and $\la^i$; thus, we can solve the relations $\dot \cH_N \approx 0$ and $\dot \cH_i \approx 0$ for $\la^N$ and $\la^i$, respectively, and compute the values of the Lagrange multipliers. Therefore, we have no further constraints, such as {\it tertiary} ones and so on. 
We will check the same for $G_A$, that the time evolution of $G_A$ defined in the phase space should also vanish on the constraint surface $\Ga_1$,
\begin{eqnarray} \label{rubu}
\dot G_A\equiv\{G_{A},\mathcal{H}_{tot}\}
&=& N\Bigg\{\frac{N}{3}\Big(2\pi-2Bp_B\Big)-2\sqrt{h}N^i\partial_iB-\sqrt{h}f''(A)\lambda^{A}\Bigg\} \non
&\approx&0 \,.
\end{eqnarray}
The role of Eq.~\eqref{rubu} is to fix the value of the Lagrange multiplier $\lambda^{A}$ as long as $f''(A)\neq0$. We demand that $f''(A) \neq 0$ so as to avoid \textit{tertiary constraints}. As a result, there are no {\it tertiary constraints} corresponding to $G_A$. The next step in our Hamiltonian analysis is to classify the constraints. 

%\subsubsection{$F(R)$ Theory constraints classification}

As shown above, we have $3$ {\it primary constraints} for $f(R)$ theory. 
They are: $$\pi_N\approx0,~~\pi_i\approx0\,,~~p_A\approx0,$$
 and $3$ {\it secondary constraints}, that are: 
 $$\mathcal{H}_{N}\approx0,~~\cH_{i}\approx0,~~G_A\approx0. $$ 
Following the definition of Poisson bracket in Eq. (\ref{poisdef}), we have: 
\begin{eqnarray}
\{\pi_N,\pi_i\}&=&\Bigg(\frac{\delta \pi_N}{\delta N_{}}\frac{\delta \pi_i}{\delta
\pi_N}-\frac{\delta \pi_N}{\delta
\pi_N}\frac{\delta \pi_i}{\delta N}\Bigg)+\Bigg(\frac{\delta
\pi_N}{\delta N^{i}_{}}\frac{\delta \pi_i}{\delta
\pi_i}-\frac{\delta \pi_N}{\delta
\pi_i}\frac{\delta \pi_i}{\delta N^{i}}\Bigg)\nonumber\\&+&\Bigg(\frac{\delta \pi_N}{\delta h_{ij}}\frac{\delta \pi_i}{\delta
\pi^{ij}}-\frac{\delta \pi_N}{\delta
\pi^{ij}}\frac{\delta \pi_i}{\delta h_{ij}}\Bigg)+\Bigg(\frac{\delta \pi_N}{\delta A}\frac{\delta \pi_i}{\delta
p_{A}}-\frac{\delta \pi_N}{\delta
p_{A}}\frac{\delta \pi_i}{\delta A}\Bigg)\nonumber\\
&+&\Bigg(\frac{\delta \pi_N}{\delta B}\frac{\delta \pi_i}{\delta
p_{B}}-\frac{\delta \pi_N}{\delta
p_{B}}\frac{\delta \pi_i}{\delta B}\Bigg)\approx0\,.
\end{eqnarray}
In a similar fashion, we can prove that: 
\begin{eqnarray}
&&\{\pi_N,\pi_N\}=\{\pi_N,\pi_i\}=\{\pi_N,p_A\}=\{\pi_N,\mathcal{H}_{N}\}=\{\pi_N,\cH_{i}\}=\{\pi_N,G_{A}\}\approx0\nonumber\\
&&\{\pi_i,\pi_i\}=\{\pi_i,p_A\}=\{\pi_i,\mathcal{H}_{N}\}=\{\pi_i,\cH_{i}\}=\{\pi_i,G_{A}\}\approx0\nonumber\\
&&\{p_A,p_A\}=\{p_A,\mathcal{H}_{N}\}=\{p_A,\cH_{i}\}\approx0\nonumber\\
&&\{\mathcal{H}_{N},\mathcal{H}_{N}\}=\{\mathcal{H}_{N},\cH_{i}\}=\{\mathcal{H}_{N},G_{A}\}\approx0\nonumber\\
&&\{\cH_{i},\cH_{i}\}=\{\mathcal{H}_{i},G_{A}\}\approx0\nonumber\\
&&\{G_{A},G_{A}\}\approx0\,.
\end{eqnarray}
 The only non-vanishing Poisson bracket on $\Ga_1$ is 
\begin{equation}
\{p_A,G_A\}=-\frac{\delta p_A}{\delta
p_{A}}\frac{\delta G_A}{\delta A}
=-\frac{\delta G_A}{\delta A}=-\sqrt{h}Nf''(A)\not\approx 0\,.
\end{equation}
Having $\{p_A,G_A\}\neq 0$ for $f''(A)\neq 0$ means that both $p_A$ and $G_A$ are \textit{second-class
constraints}.  The rest of the constraints ($\pi_N,\pi_i,\mathcal{H}_N,\mathcal{H}_{i}$) are to be counted as \textit{first-class constraints}.

%%%%%%%%%%%%%%%%%%%%%%%%%%%%%%%%%%%%%%%%

\subsubsection{Number of physical degrees of freedom for $f(R)$ gravity}

Having identified the {\it primary and secondary constraints} and categorising them into {\it first and second-class constraints}~\footnote{Having {\it first-class and second-class constraints} means there are no arbitrary functions in the Hamiltonian. Indeed, a set of canonical variables that satisfies the constraint equations determines the physical state. }, we can use the formula in Eq. (\ref{dofcount}) to count the number of the physical degrees of freedom. For $f(R)$ gravity, 
we have, 
\begin{eqnarray}
2\mathcal{A}&=&2\times\{(h_{ij}, \pi^{ij}),(N,\pi_N),(N^i, \pi_i),(A,p_A),(B,p_B)\}\nonumber\\
~&=& 2(6+1+3+1+1)=24,\nonumber\\
\mathcal{B}&=&(p_{A},G_{A})=(1+1)=2,\nonumber\\
2\mathcal{C}&=&2\times(\pi_N,\pi_i,\mathcal{H}_N,H_{i})=2(1+3+1+3)=16,\nonumber\\
\mathcal{N}&=&\frac{1}{2}(24-2-16)=3\,.
\end{eqnarray}
Hence $f(R)$ gravity has $3$ physical degrees of freedom in four dimensions; that includes the physical degrees of freedom for massless graviton and also an extra scalar degree of freedom~\footnote{We may note that the Latin indices are running from $1$ to $3$ and are spatial. Moreover, $(h_{ij}, \pi^{ij})$ pair is symmetric therefore we get $6$ from it.}.

From the Lagrangian perspective, one can study the propagator for $f(R)$ theory of gravity. The graviton propagator in such a theory can be 
computed in terms of the spin-$2$ and spin-$0$ components~\footnote{\label{oo}The propagator for $f(R)$ theory of gravity can be recast in four dimensions as, see~\cite{Biswas:2013kla}
$$\Pi(k^2) \sim \frac{1}{k^2}\left(\frac{P^{(2)}}{a}-\frac{P^{(0)}}{a-3c}\right),$$
where $P^{(2)},~P^{(0)}$ are the spin projector operators,
$a,~c$ can be functions of $k^2$. When $a=c$, we recover the propagator for Einstein-Hilbert action. For $f(R)$ gravity, $a\neq c$, which yields one extra scalar degrees of freedom. This degrees of freedom is nothing but the Brans-Dicke scalar. For details, see~\cite{Biswas:2013kla}.}.
Let us now briefly discuss few cases of interest:

\begin{itemize}

\item{Number of degrees of freedom for $f(R)=R+\alpha R^{2}$:\\
For a specific form of 
\begin{equation}
f(R)=R+\alpha R^{2}\,,
\end{equation}
where $\alpha=(6M^{2})^{-1}$ to insure correct dimensionality. In this case we have, 
\begin{equation}
\{p_A,G_A\}=-\sqrt{h}Nf''(A)=-2\sqrt{h}N\not\approx 0\,.
\end{equation}
The other Poisson brackets remain zero on the constraint surface $\Ga_1$, and hence we are left with $3$ physical degrees of freedom. }

\item{Number of degrees of freedom for $f(R)=R$:\\
For Einstein Hilbert action $f(R)$ is simply,
\begin{equation}
f(R)=R\,,
\end{equation} 
for which, 
\begin{equation}
\{p_A,G_A\}=-\sqrt{h}Nf''(A)\approx0\,.
\end{equation}
Therefore, in this case both $p_A$ and $G_A$ are {\it {first-class constraints}}. Hence, now our degrees of freedom counting formula in Eq. (\ref{dofcount})
takes the following form:
\begin{eqnarray}
2\mathcal{A}&=&2\times\{(h_{ij}, \pi^{ij}),(N,\pi_N),(N^i, \pi_i),(A,p_A),(B,p_B)\}\nonumber\\
&=&2(6+1+3+1+1)=24, \nonumber\\
\mathcal{B}&=&0,\nonumber\\
2\mathcal{C}&=&2\times(\pi_N,\pi_i,\mathcal{H}_N,H_{i},p_{A},G_{A})=2(1+3+1+3+1+1)=20,\nonumber\\
\mathcal{N}&=& \frac{1}{2}(24-0-20)=2\,,
\end{eqnarray}
which coincides with that of the spin-$2$ graviton and $a=c$ as expected from the Einstein-Hilbert action, see footnote~\ref{oo}.}
\end{itemize}

%%%%%%%%%%%%%%%%%%%%%%%%%%%%%%%%%%%%%%%%%%%%%%%%%%%%%%%%%%%%%%%%%%%

\subsection{Constraints for IDG  }

The action and the ADM decomposition of IDG has been explained explicitly in Sec \ref{idgsec}. In this section, we will
focus on the Hamiltonian analysis for the action of the form of Eq. (\ref{eq1}). The first step is to consider Eq. (\ref{finaldecomposition}),
and read off the conjugate momenta,  
\begin{eqnarray}
&&\pi_N=\frac{\p \cL}{\p \dot{N}}\approx0,\quad\pi_i=\frac{\p \cL}{\p \dot{N}^i}\approx0, \quad \pi^{ij}=\frac{\partial\mathcal{L}}{\partial\dot{h}_{ij}}=\sqrt{h}B(K^{ij}-h^{ij}K)-\sqrt{h}\nabla_\bn Bh^{ij},\nonumber\\
&&p_A=\frac{\partial\mathcal{L}}{\partial\dot{A}}=\sqrt{h}\Big[-(A\nabla_\bn \chi_1  +\chi_1
\nabla_\bn A)-\sum^{\infty}_{l=2}(\chi_l
\nabla_\bn \eta_{l-1})\Big],\quad p_B=\frac{\partial\mathcal{L}}{\partial\dot{B}}=-2\sqrt{h}K,\nonumber\\
&&p_{\chi_1}=\frac{\partial\mathcal{L}}{\partial\dot{\chi}_1}=-\sqrt{h}A\nabla_\bn A,\quad p_{\chi_l}=\frac{\partial\mathcal{L}}{\partial\dot{\chi}_l}=-\sqrt{h}(A\nabla_\bn \eta_{l-1}),\quad\nonumber\\
&&p_{\eta_{l-1}}=\frac{\partial\mathcal{L}}{\partial\dot{\eta}_{l-1}}=-\sqrt{h}(A\nabla_\bn \chi_l
+\chi_l
\nabla_\bn A).
\end{eqnarray}
as we can see in this case, the time derivatives of the lapse, i.e. $\dot N$, and the shift function, $\dot N^i$, are absent. Therefore, we have two {\it primary constraints}, 
\begin{equation}
\pi_N\approx0, \quad \pi_i\approx0\,.
\end{equation}
The total Hamiltonian is given by:
\begin{align}
H_{tot}&=\int d^{3}x \, \cH \\ \label{wowo}
&=\int d^{3}x \, \Big(N\mathcal{H}_N+N^i\cH_i+\la^{N}\pi_{N}+\la^{i}\pi_{i}\Big)\,,
\end{align}
where $\la^N$, $\la^i$ are Lagrange multipliers and the Hamiltonian density is given by: 
\begin{align}\label{Hamilton-IDG}
\mathcal{H}&=\pi^{ij}\dot{h}_{ij}+p_A\dot{A}+p_B\dot{B}+p_{\chi_1}\dot{\chi}_1+p_{\chi_l}\dot{\chi}_l+p_{\eta_{l-1}}\dot{\eta}_{l-1}-\mathcal{L}\\
&=N\mathcal{H}_N+N^i\cH_i\,,
\end{align}
using the above equation and after some algebra we have: 
\begin{eqnarray}\label{Hamilton-IDG1}
&&\mathcal{H}_{N}=\frac{1}{\sqrt{h}B}\pi^{ij}h_{ik}h_{jl}\pi^{kl}-\frac{1}{3\sqrt{h}B}\pi^2-\frac{\pi
p_B}{3\sqrt{h}}\\\nonumber&&+\frac{B}{6\sqrt{h}}p_B^{2}-\sqrt{h}BR+\sqrt{h}BA+2\partial_j[\sqrt{h}h^{ij}\partial_i]B\\\nonumber
&&-\frac{1}{A\sqrt{h}}p_{\chi_{1}}(p_A-\frac{\chi_1}{A}p_{\chi_1})-\frac{1}{A\sqrt{h}}\sum^{n}_{l=2}p_{\chi_{l}}(p_{\eta_{l-1}}-\frac{\chi_{l}}{A}p_{\chi_{1}})\\\nonumber
&&-\sqrt{h}\sum^{n}_{l=1}A\chi_l\eta_l-\sqrt{h}\frac{1}{2}A(M^{2}_{p}+f_{0}A+\sum^{\infty}_{n=1}f_n\eta_{n})\\\nonumber
&&-\sqrt{h}h^{ij}(A\partial_i\chi_1  \partial_j A+\chi_1
\partial_i A
\partial_j A)-\sqrt{h}h^{ij}\sum^{n}_{l=2}(A\partial_i\chi_l  \partial_j\eta_{l-1}
+\chi_l
\partial_i A
\partial_j \eta_{l-1})\,,
\end{eqnarray}
and, 
\begin{eqnarray}\label{Hamilton-IDG2}
\cH_i=-2h_{ik}
\nabla_l\pi^{kl}+p_A\partial_iA+p_{\chi_1}\partial_i\chi_1+p_B\partial_iB+\sum^{n}_{l=2}(p_{\chi_l}\partial_i\chi_l+p_{\eta_{l-1}}\partial_i\eta_{l-1})\,.
\nonumber\\
\end{eqnarray}
As described before in Eq. (\ref{secondaryconstraints1}), we can determine the
{\it secondary constraints},  by: 
\begin{equation}
\mathcal{H}_{N}\approx0,\quad \quad \cH_{i}\approx0\,.
\end{equation}
We can also show that, on the constraint surface $\Ga_1$, the time evolutions $\dot \cH_{N}=\{\cH_{N},\mathcal{H}_{tot}\}\approx0$ and $\dot \cH_i=\{\cH_{i},\mathcal{H}_{tot}\}\approx0$ fix the Lagrange multipliers $\la^N$ and $\la^i$,  and there will be no 
{\it tertiary constraints}.

%%%%%%%%%%%%%%%%%%%%%%%%%%%%%%%%%%%%%%%%%%%%%%%%%%%%%

\subsubsection{Classifications of constraints for IDG } 
 
 As we have explained earlier, {\it primary and secondary constrains} can be classified into {\it first or second-class constraints}. This is  derived by calculating the Poisson brackets constructed out of the constraints between themselves and each other. Vanishing Poisson brackets indicate {\it first-class constraint} and non vanishing Poisson bracket means we have {\it second-class constraint}.

 For IDG action,  we have two {\it primary constraints}: $\pi_N\approx0$ and $\pi_i\approx0$,  and two {\it secondary constraints}: $\mathcal{H}_{N}\approx0$, $H_{i}\approx0$, therefore we can determine the classification of the constraints as:
\begin{eqnarray}
\{\pi_N,\pi_i\}&=&\Bigg(\frac{\delta \pi_N}{\delta N_{}}\frac{\delta \pi_i}{\delta
\pi_N}-\frac{\delta \pi_N}{\delta
\pi_N}\frac{\delta \pi_i}{\delta N}\Bigg)+\Bigg(\frac{\delta \pi_N}{\delta N^{i}_{}}\frac{\delta \pi_i}{\delta
\pi_i}-\frac{\delta \pi_N}{\delta
\pi_i}\frac{\delta \pi_i}{\delta N^{i}}\Bigg) 
+\Bigg(\frac{\delta \pi_N}{\delta h_{ij}}\frac{\delta \pi_i}{\delta
\pi^{ij}}-\frac{\delta \pi_N}{\delta
\pi^{ij}}\frac{\delta \pi_i}{\delta h_{ij}}\Bigg)\nonumber\\
&+&\Bigg(\frac{\delta \pi_N}{\delta A}\frac{\delta \pi_i}{\delta
p_{A}}-\frac{\delta \pi_N}{\delta
p_{A}}\frac{\delta \pi_i}{\delta A}\Bigg)+\Bigg(\frac{\delta \pi_N}{\delta B}\frac{\delta \pi_i}{\delta
p_{B}}-\frac{\delta \pi_N}{\delta
p_{B}}\frac{\delta \pi_i}{\delta B}\Bigg)+\Bigg(\frac{\delta \pi_N}{\delta \chi_1}\frac{\delta \pi_i}{\delta
p_{\chi_1}}-\frac{\delta \pi_N}{\delta
p_{\chi_1}}\frac{\delta \pi_i}{\delta \chi_1}\Bigg)\nonumber\\
&+&\Bigg(\frac{\delta \pi_N}{\delta \chi_l}\frac{\delta \pi_i}{\delta
p_{\chi_l}}-\frac{\delta \pi_N}{\delta
p_{\chi_l}}\frac{\delta \pi_i}{\delta \chi_l}\Bigg)+\Bigg(\frac{\delta \pi_N}{\delta \eta_{l-1}}\frac{\delta \pi_i}{\delta
p_{\eta_{l-1}}}-\frac{\delta \pi_N}{\delta
p_{\eta_{l-1}}}\frac{\delta \pi_i}{\delta \eta_{l-1}}\Bigg)\approx 0\,.
\end{eqnarray}
In a similar manner, we can show that: 
\begin{eqnarray}
&&\{\pi_N,\pi_N\}=\{\pi_N,\pi_i\}=\{\pi_N,\mathcal{H}_{N}\}=\{\pi_N,\cH_{i}\}\approx0\nonumber\\
&&\{\pi_i,\pi_i\}=\{\pi_i,\mathcal{H}_{N}\}=\{\pi_i,\cH_{i}\}\approx0\nonumber\\
&&\{\mathcal{H}_{N},\mathcal{H}_{N}\}=\{\mathcal{H}_{N},\cH_{i}\}\approx0\nonumber\\
&&\{\cH_{i},\cH_{i}\}\approx 0\,.
\end{eqnarray}
Therefore, all of them $(\pi_N,\pi_i,\mathcal{H}_{N},\cH_{i})$ are {\it first-class constraints}. 
We can established that by solving the 
equations of motion for $\chi_n$ yields 
$$\eta_{1}=\Box A,~~\cdots,~~\eta_l=\Box \eta_{l-1}=\Box^{l}A, $$ for $l \geq 2$. Therefore, we can conclude that the $\chi_n$'s are 
Lagrange multipliers, and we get the following {\it primary constraints} from equations of motion, 
\begin{align}
\Xi_{1} = \eta_{1}-\Box A & = 0 \,, \non
\Xi_{l} = \eta_{l}-\Box \eta_{l-1}& = 0 \,,
\end{align} 
where $l \geq 2$. In fact, it is sufficient to say that $ \eta_{1}-\Box A \approx 0$ and $\eta_{l}-\Box \eta_{l-1} \approx 0$ on a constraint surface spanned by {\it primary and secondary constraints}, \textit{i.e.}, $(\pi_N\approx 0,~\pi_{i}\approx 0,~{\cal H}_{N}\approx 0,~{\cal H}_{i}\approx 0,~\Xi_{n}\approx 0)$.
As a result, we can now show, with the help of equations of motion, that we have
\begin{equation}
\{\Xi _{n},\pi_N\}=\{\Xi _{n},\pi_i\}=\{\Xi _{n},\mathcal{H}_{N}\}=\{\Xi _{n},\mathcal{H}_{i}\}=\{\Xi _{m},\Xi _{n}\}\approx 0\,;
\end{equation}
we have used, in lieu of the $=$ sign, the notation $\approx$, because the latter is a sufficient condition to be satisfied on the 
constraint surface defined by $\Ga_1=(\pi_N\approx 0,~\pi_{i}\approx 0,~{\cal H}_{N}\approx 0,~{\cal H}_{i}\approx 0,~\Xi_{n}\approx 0)$, which signifies that $\Xi _{n}$'s are now part of  {\it first-class constraints}. We should point out that we have checked that the Poisson brackets of all possible pairs among the constraints vanish on the constraint surface $\Ga_1$; as a result, there are no \textit{second-class constraints}.

%%%%%%%%%%%%%%%%%%%%%%%%%%%%%%%%%%%%%%%%%%%%%%%%%%%%%%%

\section{Physical degrees of freedom for IDG }

We can again use Eq. (\ref{dofcount}) to compute the degrees of freedom for IDG action Eq.~(\ref{eq1}). First, let us establish the
number of the phase space variables, $\mathcal{A}$. Since the auxiliary field $\chi_n$ are Lagrange multipliers, they are 
not dynamical and hence redundant, as we have mentioned earlier. In contrast we have to count the $(B,~p_b)$ pair in 
the phase space as $B$ contains intrinsic value. For the IDG action Eq.~(\ref{eq1}), we have:
\begin{eqnarray}\label{phasespacevariables}
2\mathcal{A}&\equiv&2\times\bigg\{(h_{ij}, \pi^{ij}),(N,\pi_N),(N^i,
\pi_i),(B,p_b),(A,p_A),\underbrace{(\eta_1,p_{\eta_1}),(\eta_2,p_{\eta_2}),\cdots}_{n=1,~2,~3,\cdots \infty}\bigg\}\nonumber\\
&\equiv&2\times(6+1+3+1+1+\infty)=24+\infty\,,
\end{eqnarray}   
trivially we have $(\eta_n,~p_{\eta_n})$ and for each pair we have assigned one variable, which is multiplied by a factor of $2$ since we are dealing with field-conjugate momentum pairs in the phase space. Moreover, as we have found from the Poisson brackets of all possible pairs among the constraints, the number of the {\it second-class constraints}, $\mathcal{B}$, is equal to zero. In the next sub-sections, we will show that the \textit{correct} number of the {\it first-class constraints} depends on the choice of $\mathcal{F}(\Box)$.

%%%%%%%%%%%%%%%%%%%%%%%%%%%%%%%%%%%%%%%%%%%%%%%%%%%

\subsection{Choice of $\mathcal{F}(\bbox)$}
\label{mimi}

In this section, we will focus on an appropriate choice of ${\cal F}(\Box)$ for the action Eq.(\ref{eq1}). From the Lagrangian point of view, we could analyse the propagator of the action Eq.(\ref{eq1}). It was found in Refs.~\cite{Biswas:2005qr,Biswas:2011ar} that $\mathcal{F}(\bbox)$ can take the following form,
\begin{equation}\label{generalfbox}
\mathcal{F}(\bbox)=M^{2}_{p}\frac{c(\bbox)-1}{\Box}\,.
\end{equation}
The  choice of $c(\bbox)$ determines how many roots we have and how many poles are present in the graviton propagator, see Refs.~\cite{Biswas:2005qr,Biswas:2011ar,Biswas:2013kla}. Here, we will consider two choices of $c(\bbox)$, one which has infinitely many roots, and therefore infinite poles in the propagator. 
 For instance, we can choose 
\begin{equation}
c(\bbox)=\cos(\bbox)\,,
\end{equation}
then the equivalent action would be written as: 
\begin{equation}\label{badaction}
S_{eqv}=\frac{1}{2}\int d^{4} x \, \sqrt{-g}\bigg[M^{2}_{p}\bigg(A+A\big(\frac{\cos(\bbox)-1}{\Box}\big)A\bigg
)+B(R-A)\bigg]\,.
\end{equation}
By solving the equations of motion for $A$, and subsequently solving for $\cos(\bbox)$
we get,
\begin{eqnarray}
\cos(\bar{k}^{2})=1-\frac{k^{2}(BM^{-2}_{p}-1)}{2A}\,,
\end{eqnarray}
where in the momentum space, we have $(\Box\rightarrow -k^2)$, and also
 note $\bar{k}\equiv k/M$; where
$B$ has mass dimension $2$. From~\eqref{lklk} in appendix~\ref{appendixc}, we have that
\be
B=M^{2}_{p}\left(1+\frac{4A}{3k^{2}}\right) \,.
\ee
Therefore, solving $\cos (\kb^2)=\frac{1}{3}$, we obtain infinitely many solutions. We observe that there is an infinite number of solutions; hence, there are also infinitely many degrees of freedom. Thus, we have got infinitely many solutions, which can be written schematically as:\  
\begin{align}\label{qwqw}
&\Psi_{1}=\Box A+a_1 A=0\,,\nonumber\\
&\Psi_{2}=\Box A+a_2 A=0\,,\nonumber\\
&\Psi_{3}=\Box A +a_{3}A=0\,,\nonumber\\
&\vdots
\end{align}
or, in the momentum space,
\begin{align}
-A k^{2}+Aa_1 =0 & \Ra k^{2}=a_{1} \,,\nonumber\\
-Ak^{2}+Aa_2 =0 & \Ra k^{2}=a_{2} \,,\nonumber\\
-Ak^{2} +Aa_3=0 & \Ra k^{2}=a_{3} \,,\nonumber\\
&\vdots
\end{align}
Now, acting the $\Box$ operators on Eq.~\eqref{qwqw}, we can write
\begin{align}
\Box \Psi_{2}&=\Box^{2}A+a_{2}\Box A \,, \non
\Box^{2} \Psi_{3}&=\Box^{3}A+a_{3}\Box^{2} A \,, \non
\vdots \non
\Box^{n-1} \Psi_{n}&=\Box^{n}A+a_{n}\Box^{n-1} A \,, \non
\vdots
\end{align}
Following the prescription in section ~\ref{idgsec}, we can parameterize the terms of the form $\Box A$, $\Box^{2}A$, etc. by employing the auxiliary fields $\chi_l$,  $\eta_{l}$, for $l \geq 1$. 
Therefore, we can write the solutions $\Psi_n$ as follows:
\begin{align}
&\Psi_{1}^{'}=\eta_{1}+a_1 A=0\,,\nonumber\\
&\Psi_{2}^{'}=\eta_{2}+a_2 \eta_{1}=0  \,,\nonumber\\
&\Psi_{3}^{'}=\eta_{3} +a_{3}\eta_{2}=0  \,.\nonumber\\
&\vdots
\end{align}
We should point out that we have acted the operator $\Box$ on $\Psi_{2}$, the operator $\Box^2$ on $\Psi_3$, etc. in order to obtain $\Psi_{2}^{'}$, $\Psi_{3}^{'}$, etc. As a result, we can rewrite the term $A+M_{p}^{-2}A \cF(\bbox) A$, as
\be
A+M_{p}^{-2}A \cF(\bbox)A=a_{0}\Psi_{1}^{'}\prod_{n=2}^{\infty}\Box^{-n+1}\Psi_{n}^{'} \,.
\ee
We would also require $\phi_n$ auxiliary fields acting like Lagrange multipliers, along with $\psi_{n}$ variables. Now, absorbing the powers of $M^{-2}$ into the coefficients where appropriate, 
\begin{align}
S_{eqv}&=\frac{1}{2}\int d^{4} x \, \sqrt{-g}\bigg[M^{2}_{p}a_{0}\prod_{n=1}^{\infty}\psi_{n}+B(R-A)+\chi_{1}A(\eta_{1}-\Box A)\non
&+\sum
^{\infty}_{l=2}\chi_lA(\eta_l-\Box \eta_{l-1})+\phi_{1}(\psi_{1}-\Psi_{1}^{'})+\sum^{\infty}_{n=2}\phi_n\big(\psi_n-\Box^{-n+1}\Psi_{n}^{'}\big)\bigg]\,,
\end{align}
where  $a_0$ is a constant and, let us define $\Phi_{1}=\psi_{1}-\Psi_{1}^{'}$ and, for $n \geq 2$, $\Phi_n=\psi_n-\Box^{-n+1}\Psi_n^{'}$. 
Then the equations of motion for $\phi_n$ will yield:
\begin{equation}\label{molo}
\Phi_n=\psi_n-\Box^{-n+1}\Psi_n^{'}=0\,.
\end{equation}
Again, it is sufficient to replace $\psi_n-\Box^{-n+1}\Psi_n^{'}=0$ with $\psi_n-\Box^{-n+1}\Psi_n^{'}\approx 0$ satisfied at the constraint surface.
As a result there are $n$ {\it primary constraints} in $\Phi_n$.  Moreover, by taking the equations of motion for $\chi_{n}$'s and $\phi_{n}$'s simultaneously, we will
obtain the {\it original} action, see Eq.~\eqref{badaction}. The time evolutions of the $\Xi_n$'s \& $\Phi_n$'s fix the corresponding Lagrange multipliers $\la^{\Xi_n}$ \& $\la^{\Phi_n}$ in the total Hamiltonian (when we add the terms $\la^{\Xi_n}\Xi_{n}$ \& $\la^{\Phi_n}\Phi_{n}$ to the integrand in~\eqref{wowo}); hence, the $\Xi_n$'s \& $\Phi_n$'s do not induce \textit{secondary constraints}.

Now, to classify these constraints, we can show that the following Poisson brackets involving $\Phi_n$ on the constraint surface $(\pi_N\approx0,\pi_i\approx0,\mathcal{H}_N\approx0,\mathcal{H}_{i}\approx0,\Xi_{n}\approx0,\Phi_n\approx0)$ are satisfied~
\footnote{ Let us note again that $\Ga_1$ is a smooth submanifold of the phase space determined by the {\it primary} and  {\it secondary constraints}; hereafter in this section, we shall exclusively use the ``$\approx$'' notation to denote equality on $\Ga_1$.}:
\begin{equation}
\{\Phi_n,\pi_N\}=\{\Phi_n,\pi_i\}=\{\Phi_n,\mathcal{H}_{N}\}=\{\Phi_n,\mathcal{H}_{i}\}=\{\Phi_m,\Xi _{n}\}=\{\Phi_m,\Phi_n\}\approx0\,,
\end{equation}
which means that the $\Phi_n$'s can be treated as {\it first-class constraints}. We should point out that we have checked that the Poisson brackets of all possible pairs among the constraints vanish on the constraint surface $\Ga_1$; as a result, there are no \textit{second-class constraints}. Now, from Eq.~\eqref{phasespacevariables}, we obtain:
\begin{eqnarray}
&&2\mathcal{A}\equiv2\times\bigg\{(h_{ij}, \pi^{ij}),(N,\pi_N),(N^i,
\pi_i),(B,p_b),(A,p_A),\underbrace{(\eta_1,p_{\eta_1}),(\eta_2,p_{\eta_2}),\cdots}_{n}\bigg\}
\nonumber\\&&=2\times(6+1+3+1+1+\infty)=24+\infty\nonumber\\
&&\mathcal{B}=0,\nonumber\\
&&2\mathcal{C}\equiv2\times(\pi_N,\pi_i,\mathcal{H}_N,\mathcal{H}_{i},\Xi
_{n},\Phi_n)=2(1+3+1+3+\infty+\infty)=16+\infty+\infty,\nonumber\\
&&\mathcal{N}=\frac{1}{2}(2\mathcal{A}-\mathcal{B}-2\mathcal{C})=\infty\,.
\end{eqnarray}
As we can see a bad choice for $\mathcal{F}(\bbox)$ can lead to infinite number of degrees of freedom., and there are many such examples. However, our aim is to
come up with a concrete example where IDG will be determined solely by massless graviton and at best one massive scalar in the context of Eq.~(\ref{eq1}).

%%%%%%%%%%%%%%%%%%%%%%%%%%%%%%%%%%%%%%%%%%%%%%%%

\subsection{ ${\cal F}(e^{\bbox}) $}

In the definition of $\mathcal{F}(\bbox)$ as given in Eq. (\ref{generalfbox}), if 
\begin{equation}
c(\bbox)= e^{-\gamma(\bbox)},
\end{equation}
where $\gamma (\bbox)$ is an entire function, we can decompose the propagator into partial fractions and have just one extra pole apart from the spin-2 graviton. Consequently, in order to have just one extra degree of freedom, we have to impose conditions on the coefficient in $\mathcal{F}(\bbox)$ series expansion. Moreover, to avoid $\Box^{-1}$ terms appearing in the $\mathcal{F}(\bbox)$, we must have that, 
\begin{equation}
c(\bbox)=\sum^{\infty}_{n=0}c_n\bbox^n\,,
\end{equation}
with the first coefficient $c_0=1$, therefore: 
\begin{equation}
\mathcal{F}(\bbox)=\Big(\frac{M_p}{M}\Big)^{2}\sum^{\infty}_{n=0}c_{{n+1}}\bbox^{n}\,,
\end{equation}
Suppose we have $c(\bbox)=e^{-\bbox}$, then using Eq. (\ref{generalfbox}) we have, 
 \begin{equation}
\mathcal{F}(\bbox)=\sum^{\infty}_{n=0}f_n\bbox^n\,,
\end{equation}
where the coefficient $f_n$ has the form of, 
\begin{equation}
f_n=\Big(\frac{M_p}{M}\Big)^{2}\frac{(-1)^{n+1}}{(n+1)!}\,,
\end{equation}
Indeed this particular choice of $c(\bbox)$ is very well motivated from string field theory~\cite{Biswas:2005qr}. In fact the above choice of $\gamma(\bbox )=-\bbox$ contains at most one extra zero in the propagator corresponding to one extra scalar mode in the spin-0 component of the graviton propagator~\cite{Biswas:2011ar,Biswas:2013kla}. We rewrite the action as: 
\begin{equation}\label{goodaction}
S_{eqv}=\frac{1}{2}\int d^{4} x \, \sqrt{-g}\bigg[M^{2}_{p}\bigg(A+A\big(\frac{e^{-\bbox}-1}{\Box}\big)A\bigg )+B(R-A)\bigg]\,.
\end{equation}
The equation of motion for $A$ is then:
\begin{equation}
M^{2}_{p}\bigg(1+2\big(\frac{e^{-\bbox}-1}{\Box}\big)A\bigg
)-B=0\,.
\end{equation}
In momentum space, we can solve the equation above:
\begin{eqnarray}\label{sol-for-k}
e^{\bar{k}^{2}}=1-\frac{k^{2}(BM^{-2}_{p}-1)}{2A}\,,
\end{eqnarray}
where  in the momentum space $\Box\rightarrow -k^2$ and also $\bar{k}\equiv k/M$. From Eq.~\eqref{toko} in the appendix~\ref{appendixc}, we have, $e^{\kb^2}=\frac{1}{3}$, therefore solving Eq.~(\ref{sol-for-k}), we obtain
\be
B=M^{2}_{p}\left(1+\frac{4A}{3k^{2}}\right) \,.
\ee
Note that we obtain only one extra solution (apart from the one for the massless spin-$2$ graviton). We observe that there is a finite number of real solutions; hence, there are also finitely many degrees of freedom. The form of the solution can be written schematically, as: 
\begin{equation}\label{Sol-IDG}
\Omega=\Box A+b_1A=0\,,
\end{equation}
or, in the momentum space,
\be
-Ak^{2}+Ab_{1}=0 \Ra k^{2}=b_{1} \,,
\ee
Now, following again the prescription laid down in section~\ref{idgsec}, we can parameterize the terms like $\Box A$, $\Box^{2}A$, etc. with the help of auxiliary fields $\chi_l$ and $\eta_{l}$, for $l \geq 1$. Therefore, equivalently,
\be
\Omega^{'}=\eta_{1}+b_{1}A=0 \,.
\ee
Consequently, we can also rewrite the term $A \cF(\bbox) A$ with the help of auxiliary fields $\rho$ and $\omega$. Upon taking the equations of motion for the field $\rho$, one can recast $A+M_{p}^{-2}A \cF(\bbox)A=b_{0}\omega~{\cal G}(A,\eta_{1},\eta_{2},\dots)$. Hence, we  can recast the action, Eq. (\ref{goodaction}), as, 
\begin{align}
S_{eqv}&=\frac{1}{2}\int d^{4} x \, \sqrt{-g}\bigg[M^{2}_{p}b_{0}\omega~{\cal G}(A,\eta_{1},\eta_{2},\dots)+B(R-A)+\chi_{1}A(\eta_{1}-\Box A)\non
&+\sum
^{\infty}_{l=2}\chi_lA(\eta_l-\Box
\eta_{l-1})+\rho\big(\omega-\Omega^{'}\big)\bigg]\,,
\end{align} 
where $b_0$ is a constant,  and we can now take $\rho$ as a Lagrange multiplier. The equation of motion for $\rho$ will yield:
\begin{equation}\label{pppp}
\Theta=\omega-\Omega^{'}=0\,.
\end{equation}
Note that $\Theta =\omega-\Omega^{'}\approx 0$ will suffice on the constraint surface determined by {\it primary and secondary constraints}
$(\pi_N\approx0,\pi_i\approx0,\mathcal{H}_N\approx0,\mathcal{H}_{i}\approx0,\Xi_{n}\approx0,\Theta\approx0)$. As a result, $\Theta$ is a {\it primary constraint}. 
The time evolutions of the $\Xi_n$'s \& $\Theta$ fix the corresponding Lagrange multipliers $\la^{\Xi_n}$ \& $\la^{\Theta}$ in the total Hamiltonian (when we add the terms $\la^{\Xi_n}\Xi_{n}$ \& $\la^{\Theta}\Theta$ to the integrand in~\eqref{wowo}); hence, the $\Xi_n$'s \& $\Theta$ do not induce \textit{secondary constraints}.

Furthermore, the function ${\cal G}(A,\eta_{1},\eta_{2},\dots)$ contains the root corresponding to the massless spin-$2$ graviton. Furthermore, taking the equations of motion for $\chi_{n}$'s and $\rho$ simultaneously yields the same equation of motion as that of in Eq.~\eqref{goodaction}. The
Poisson bracket of $\Theta$ with other constraints will give rise to
\begin{equation}
\{\Theta,\pi_N\}=\{\Theta,\pi_i\}=\{\Theta,\mathcal{H}_{N}\}=\{\Theta,\mathcal{H}_{i}\}=\{\Theta,\Xi
_{n}\}=\{\Theta,\Theta\}\approx0\,,
\end{equation}
where $\approx$ would have been sufficient. This leads to $\Theta$ as a {\it first-class constraint}. Hence, we can calculate the number of the physical degrees of freedom  as: 
\begin{eqnarray}
&&2\mathcal{A}\equiv2\times\bigg\{(h_{ij}, \pi^{ij}),(N,\pi_N),(N^i,
\pi_i),(B,p_b),(A,p_A),\underbrace{(\eta_1,p_{\eta_1}),(\eta_2,p_{\eta_2}),\cdots}_{n}\bigg\}
\nonumber\\&&=2\times(6+1+3+1+1+\infty)=24+\infty\nonumber\\
&&\mathcal{B}=0,\nonumber\\
&&2\mathcal{C}\equiv2\times(\pi_N,\pi_i,\mathcal{H}_N,\mathcal{H}_{i},\Xi
_{n},\Theta)=2(1+3+1+3+\infty+1)=18+\infty,\nonumber\\
&&\mathcal{N}=\frac{1}{2}(2\mathcal{A}-\mathcal{B}-2\mathcal{C})=\frac{1}{2}(24+\infty-0-18-\infty)=3\,.
\end{eqnarray}
This gives $2$ degrees of freedom from the massless spin-$2$ graviton in addition to an extra degree of freedom as expected from the propagator analysis; see Appendix \ref{appendixc}.

%%%%%%%%%%%%%%%%%%%%%%%%%%%%%%%%%%%%%%%%

\section{Conclusion}
\numberwithin{equation}{section}
\label{sec:conc}

The results of the paper support the original idea that both Lagrangian and Hamiltonian analysis will yield similar conclusions for infinite derivative theories with Gaussian kinetic term~\cite{Biswas:2005qr}. Here we have shown the dynamical resemblance explicitly by studying the degrees of freedom from 
Hamiltonian constraints. It has been known that from Lagrangian perspective the dynamical degrees of freedom can be studied from the propagator, especially from the number of poles appearing in the propagator, whether they are finite or infinite. In case of IDG, one can study the 
scalar and the tensor components of the propagating degrees of freedom~\cite{Biswas:2011ar,Biswas:2013kla}, and for Gaussian kinetic term which determines
 ${\cal F}(\bbox)$, there are only $2$ dynamical degrees of freedom. The key lesson is to make sure that there are no poles other than the original poles (corresponding to the original degrees of freedom ) in the propagator, this can be achieved by 
demanding that the propagator be suppressed by {\it exponential of an entire function}. An entire function does not have any ploes in the finite complex plane, it has essential singularities in the boundary, in the limit when $\bbox \rightarrow \infty$. This choice of propagator determines the kinetic term in Lagrangian for infinite derivative theories. For a scalar toy model the kinetic term becomes Gaussian, \textit{i.e.}, ${\cal F}=\Box e^{-\bbox}$, while in gravity it becomes $ {\cal F}=M_{p}^{2}\Box^{-1}(e^{-\bbox}-1)$.
 
 From the Hamiltonian perspective, the essence of finding the dynamical degrees of freedom relies primarily on finding the total phase space variables, and {\it first} and {\it second-class constraints}. As expected, infinite derivative theories  will have infinitely many phase space variables, and so will be {\it first} and {\it second-class constraints}. However,  for a Gaussian kinetic term, ${\cal F}(\bbox)$, the degrees of freedom are indeed finite. We show this for both scalar and  gravitational Hamiltonian densities. In the case of gravity, seeking the Hamiltonian density requires a careful handling due to diffeomorphism invariance involved in temporal evolution. In this paper, we present the Hamiltonian density for IDG~\cite{Biswas:2005qr}, which contains infinite derivatives only in the Ricci scalar. We will provide the full Hamiltonian density for the full quadratic curvature gravity in future publication, which will involve Ricci curvature and the Weyl term.

\section*{Acknowledgments}

ST is supported by a scholarship from the Onassis Foundation. 
\appendix

%%%%%%%%%%%%%%%%%%%%%%%%%%%%%%%%%%%%%%%%%%%%%%%

\section{Hamiltonian density}\label{hamdensapp}
Hamiltonian density corresponding to action Eq.~(\ref{hamdens}) is explicitly given by, 
\begin{eqnarray}\label{hamapp-1}
\mathcal{H}&=&p_A\dot{A}+p_{\chi_1}\dot{\chi}_1+p_{\chi_l}\dot{\chi}_l+p_{\eta_{l-1}}\dot{\eta}_{l-1}-\mathcal{L}\nonumber\\
&=&-(A\dot{\chi_1}
 \dot{A}+\dot{A}\chi_1
\dot{A})-\sum^{\infty}_{l=2}(\dot{A}\chi_l\dot{\eta}_{l-1})\nonumber\\
&-&(A\dot{
A)}\dot{\chi}_1-(A\dot{\eta}_{l-1})\dot{\chi}_l-(A\dot{\chi}_l
\dot{\eta}_{l-1}+\chi_l
 \dot{A}\dot{\eta}_{l-1})\nonumber\\
 &-&\Bigg(A(f_{0}A+\sum^{\infty}_{n=1}f_n\eta_{n})+\sum ^{\infty}_{l=1}A\chi_l
\eta_l\nonumber\\
&-&(A\partial_0\chi_1  \partial_0 A+\chi_1
\partial_0 A
\partial_0 A)+\eta^{ij}(A\partial_i\chi_1  \partial_j A+\chi_1
\partial_i A
\partial_j A)\nonumber\\
&-&\sum^{\infty}_{l=2}(A\partial_0\chi_1  \partial_0\eta_{l-1}
+\chi_l
\partial_0 A
\partial_0 \eta_{l-1})+\eta^{ij}\sum^{\infty}_{l=2}(A\partial_i\chi_l  \partial_j\eta_{l-1}
+\chi_l
\partial_i A
\partial_j \eta_{l-1})\Bigg)\nonumber\\
&=&-\sum^{\infty}_{l=2}(\dot{A}\chi_l\dot{\eta}_{l-1})
-(A\dot{
A)}\dot{\chi}_1-(A\dot{\eta}_{l-1})\dot{\chi}_l
 +\Bigg(A(f_{0}A+\sum^{\infty}_{n=1}f_n\eta_{n})-\sum ^{\infty}_{l=1}A\chi_l
\eta_l\nonumber\\ \\
&-&\eta^{ij}(A\partial_i\chi_1  \partial_j A+\chi_1
\partial_i A
\partial_j A)
-\eta^{ij}\sum^{\infty}_{l=2}(A\partial_i\chi_l  \partial_j\eta_{l-1}
+\chi_l
\partial_i A
\partial_j \eta_{l-1})\Bigg)\nonumber\\ \\
&=&
A(f_{0}A+\sum^{\infty}_{n=1}f_n\eta_{n})-\sum ^{\infty}_{l=1}A\chi_l
\eta_l\nonumber\\
&-&(\eta ^{\mu\nu}A\partial_{\mu}\chi_1  \partial_\nu A+g^{ij}\chi_1
\partial_i A
\partial_j A)
-\eta^{\mu\nu}\sum^{\infty}_{l=2}(A\partial_\mu\chi_l  \partial_\nu\eta_{l-1}
+\chi_l
\partial_\mu A
\partial_\nu \eta_{l-1})\,.\nonumber\\
\end{eqnarray}

%%%%%%%%%%%%%%%%%%%%%%%%%%%%%%%%%%%%%%%%%%%%%%%%%%%%%%%%%%%%%%%%

\section{Auxiliary fields $\chi_1$ and $\eta_1$}\label{app1}
The right-hand side of Eq.(\ref{additionalterm}) can be derived as: 
\begin{eqnarray}
A(\Box A)&\Rightarrow&\chi_1 A(\eta_1-\Box A)=\chi_1 A\eta_1-\chi_1 A\Box A \nonumber\\
&=&\chi_1 A\eta_1-g^{\mu\nu}\chi_1 A\partial_\mu \partial_\nu A\nonumber\\&=&\chi_1 A\eta_1-g^{\mu\nu}\partial_\mu(\chi_1
A \partial_\nu A)+g^{\mu\nu}\partial_\mu\chi_1 A \partial_\nu A+g^{\mu\nu}\chi_1
\partial_\mu A
\partial_\nu A\nonumber\\
&=&\chi_1 A\eta_1+g^{\mu\nu}\partial_\mu\chi_1 A \partial_\nu A+g^{\mu\nu}\chi_1
\partial_\mu A
\partial_\nu A \,,
\end{eqnarray}
where it should be noted that we have dropped the total derivative and also we have absorbed the factor of $M^{-2}$ into $c_1$ \& $\chi_1$ (the mass dimension of $\eta_1$ is modified accordingly), and, hence, here the d'Alembertian operator is not barred.
  
\section{$B(R-A)$ decomposition}\label{app2}
We can decompose the $B(R-A)$ as: 
\begin{eqnarray}
B(R-A)
&=&B\Big(K_{ij}K^{ij}-K^2+\mathcal{R}-A\Big)+B\frac{2}{\sqrt{h}}\partial_\mu(\sqrt{h}n^\mu
K)\nonumber\\&-&B\frac{2}{N\sqrt{h}}\partial_i(\sqrt{h}h^{ij}\partial_j N)\,,
\end{eqnarray}
where: 
\begin{eqnarray}
&&B\frac{2}{\sqrt{h}}\partial_\mu(\sqrt{h}n^\mu
K)\\\nonumber &&=\nabla_\mu\Big[B\frac{2}{\sqrt{h}}(\sqrt{h}n^\mu
K)\Big]-(\nabla_\mu B)\frac{2}{\sqrt{h}}(\sqrt{h}n^\mu
K)\\\nonumber &&=\nabla_\mu\Big[2Bn^\mu
K\Big]-2(\nabla_\mu B)n^\mu
K=-2n^\mu(\nabla_\mu B)K=-2\nabla_\bn BK\,,
\end{eqnarray}
and,
\begin{eqnarray}
&&-B\frac{2}{N\sqrt{h}}\partial_i(\sqrt{h}h^{ij}\partial_j N)\\ \nonumber&&=-\frac{2}{N\sqrt{h}}\partial_i(B(\sqrt{h}h^{ij}\partial_j
N))+\frac{2}{N\sqrt{h}}\partial_i(B)\sqrt{h}h^{ij}\partial_j
N\\\nonumber
&&=\frac{2}{N\sqrt{h}}\partial_i(B)\sqrt{h}h^{ij}\partial_j
N=\frac{2}{N\sqrt{h}}\partial_j(\partial_i(B)\sqrt{h}h^{ij}
N)-\frac{2}{\sqrt{h}}\partial_j(\partial_i(B)\sqrt{h}h^{ij}
)\\\nonumber
&&=-\frac{2}{\sqrt{h}}\partial_j(\partial_i(B)\sqrt{h}h^{ij}
)\,.
\end{eqnarray}

%%%%%%%%%%%%%%%%%%%%%%%%%%%%%%%%%%%%%%%%%%%%%%%%%%%%%

\section{Finding the  physical degrees of freedom from propagator
analysis}\label{appendixc}

We have an action of the form~\cite{Biswas:2005qr}
\be
S=\frac{1}{2} \int d^{4}x \, \sqrt{-g} \LT M_{P}^{2}R +R\cF\left(\bbox\right)R \RT
\ee
or, equivalently,
\be \label{luwo}
S=\frac{1}{2} \int d^{4}x \, \sqrt{-g} \LT M_{P}^{2}A +A\cF\left( \bbox
\right)A +B(R-A)\RT \,.
\ee
$A$ and $B$ have mass dimension $2$.

The propagator around Minkowski spacetime is of the form~\cite{Biswas:2011ar,Biswas:2013kla}
\be
\Pi (-k^2)=\frac{\cP ^ 2}{k^{2}a(-k^2)}+\frac{\cP_{s} ^ 0}{k^{2}(a(-k^2)-3c(-k^2))} \,,
\ee
where $a(\Box)=1$ and $c(\Box)=1+M_{P}^{-2}\cF \left( \bbox \right)\Box$.
Hence,
\be \label{ol}
\Pi (-k^2)=\frac{\cP ^ 2}{k^{2}}+\frac{\cP_{s} ^ 0}{k^{2}(-2+3M_{P}^{-2}k^{2}\cF(-k^{2}/M^{2}))}
\ee
We know that~\cite{Biswas:2011ar,Biswas:2013kla}
\be
\cF (\bar{\Box})=M_{P}^{2}\frac{c(\bar{\Box})-1}{\Box} \,.
\ee
Only if $c(\Box)$ is the exponent of an entire function can we decompose into partial fractions and have just one extra pole.

The upshot is that, in order to have just one extra degree of freedom, we have to impose conditions on the coefficients in $\cF \LF \bbox \RF$. In order to avoid $\Box^{-1}$ terms appearing in $\cF{(\bar{\Box})}$, we must have that
\be
c(\bar{\Box})=\sum_{n=0}^{\infty}c_{n}\bar{\Box}^{n} 
\ee
and $c_{0}=1$. Hence,
\be
\cF(\bar{\Box})=\LF \frac{M_{P}}{M} \RF^{2}\sum_{n=0}^{\infty}c_{n+1}\bar{\Box}^{n} \,.
\ee
To get infinitely many poles and, hence, degrees of freedom, one could have, for instance, that
\be
c(\bar{\Box})= \cos (\bar{\Box}) \,,
\ee
so that $c_{0}=1$. Then Eq.~\eqref{luwo} becomes
\be 
S=\frac{1}{2} \int d^{4}x \, \sqrt{-g} \LT M_{P}^{2}A +M_{P}^{2}A\left(\frac{\cos (\bar{\Box})-1}{\Box}\right)A +B(R-A)\RT \,.
\ee
Using~\eqref{ol}, apart from the $k^2=0$ pole,  we have poles when
\be \label{lklk}
\cos \left(\frac{k^{2}}{M^2}\right)=\frac{1}{3}\,.
\ee 
Eq.~\eqref{lklk} has infinitely many solutions due to the periodicity of the cosine function and, therefore, the propagator has infinitely many poles and, hence, degrees of freedom.
We can
write
the solutions as $\bar{k}^{2}=2m\pi$, where $m=0,1,2,\cdots$, one can also
write: 
\begin{equation}
\cos(\bar{k}^{2})=\prod^{\infty}_{l=1}\Bigg(1-\frac{4\bar{k}^{4}}{(2l-1)^{2}\pi^{2}}\Bigg)
\end{equation}
or 
\begin{equation}
\cos(\bbox)=\prod^{\infty}_{l=1}\Bigg(1-\frac{4\bbox^{2}}{(2l-1)^{2}\pi^{2}}\Bigg)
\end{equation}
Now, to get just one extra degrees of freedom, one can make, for instance, the choice $c(\bar{\Box})=e^{-\bar{\Box}}$, then
\be
\cF(\bar{\Box})=\sum_{n=0}^{\infty}f_{n}\bar{\Box}^{n} \,,
\ee
where
\be
f_{n}=\LF \frac{M_P}{M} \RF^{2}\frac{(-1)^{n+1}}{(n+1)!} \,.
\ee
Using~\eqref{ol}, apart from the $k^2=0$ pole, we have poles when
\be \label{toko}
e^{k^{2}/M^{2}}=\frac{1}{3}\,.
\ee
There is just one extra pole and, hence, degrees of freedom. In total, there are $3$ degrees of freedom.

%+Bibliography

%-Bibliography

\end{document}